\documentstyle[prd,aps,epsfig]{revtex}

\newcommand{\wpr}{$W^\prime$~}
\newcommand{\wprm}{W^\prime}
\newcommand{\wprmp}{W^{\prime\,+}}
\newcommand{\mwpsq}{m^2_{W^\prime}}
\newcommand{\mtsq}{m^2_{t}}

\newcommand{\beq}{\begin{equation}}
\newcommand{\eeq}{\end{equation}}
\newcommand{\bea}{\begin{eqnarray}}
\newcommand{\eea}{\end{eqnarray}}

\begin{document}

\preprint{FERMILAB-Pub-02/165-T}

\title{Fully differential $W^\prime$ production and decay at next-to-leading
order in QCD}

\author{Zack Sullivan}
\address{Theoretical Physics Department, Fermi National
Accelerator Laboratory, Batavia, IL, 60510-0500}

\date{July 23, 2002}

\maketitle

\begin{abstract}
We present the fully differential production and decay of a $W^\prime$
boson, with arbitrary vector and axial-vector couplings, to any final
state at next-to-leading order in QCD.  We demonstrate a complete
factorization of couplings at next-to-leading order in both the
partial width of the $W^\prime$ boson, and in the full two-to-two
cross section.  We provide numerical predictions for the contribution
of a $W^\prime$ boson to single-top-quark production, and separate
results based on whether the mass of the right-handed neutrino $\nu_R$
is light enough for the leptonic decay channel to be open.  The
single-top-quark analysis will allow for an improved direct $W^\prime$
mass limit of 525--550~GeV using data from run I of the Fermilab
Tevatron.  We propose a modified tolerance method for estimating
parton distribution function uncertainties in cross sections.
\end{abstract}

\section{Introduction}

Since the introduction of the standard model, many extensions have
been proposed that involve enhanced gauge symmetries.  One common
feature of these models is the prediction of additional gauge bosons,
generically called \wpr and $Z^\prime$ bosons.  In non-universal and
top-flavor models, the \wpr gauge boson arises from a new SU(2)$_L$
sector that distinguishes between generations of fermions
\cite{Li:nk,Malkawi:1996fs,Muller:1996dj}, and may treat quarks and
leptons differently \cite{Georgi:1989ic,Georgi:1989xz,He:1999vp}.  The
\wpr boson could be the lowest Kaluza-Klein mode of the $W$ boson
\cite{Datta:2000gm}, or a heavy mass eigenstate in non-commuting
extended technicolor \cite{Chivukula:1995gu}.  In a left-right
symmetric model the \wpr and standard-model $W$ bosons are remnants of
a broken ${\mathrm SU(2)}_L \times {\mathrm SU(2)}_R$ symmetry
\cite{Pati:1974yy,Mohapatra:1974hk,Mohapatra:1974gc,Senjanovic:1975rk}.
The resulting left-right mixing may be naturally suppressed by
orbifold breaking of the left-right symmetry \cite{Mimura:2002te}, or
by supersymmetric interactions \cite{Cvetic:1983su}.  While indirect
bounds may be placed on the masses and couplings of these various \wpr
bosons within the context of their explicit theories, it is always
advantageous to search for these new particles directly.  In order to
facilitate a direct experimental search, and to achieve the most
general contact with theory, we calculate the fully differential
next-to-leading order cross section for the production and decay of a
\wpr boson to any final state with arbitrary vector and axial-vector
couplings.

The most general Lorentz invariant Lagrangian describing the coupling
of a \wpr to fermions may be written as
\beq
{\cal L} = \frac{g}{\sqrt{2}} \overline{f}_i\gamma_\mu \bigl(
C^R_{f_if_j} P_R + C^L_{f_if_j} P_L \bigr) W^{\prime} f_j +
\mathrm{H.c.} \,, \label{eq:genlan}
\eeq
where $P_{R,L}=(1\pm \gamma_5)/2$, $g$ is the standard model SU(2)$_L$
coupling, and the $C^{R,L}_{f_if_j}$ are arbitrary couplings that
differ for quarks and leptons.  For a standard model $W$ boson,
$C^R=0$, and $C^L$ is either the Cabibbo-Kobayashi-Maskawa matrix or
diagonal, for quarks or leptons, respectively.

Most experimental searches have concentrated on \wpr bosons that decay
via a purely left- or right-handed current.  In order to make contact
with these results, we rewrite Eq.~(\ref{eq:genlan}) in the notation
typical of left-right symmetric models as \cite{Langacker:1989xa}
\beq
{\cal L} = \frac{1}{\sqrt{2}} \overline{f}_i\gamma_\mu \bigl( g_R
e^{i\omega} \cos\zeta\, V^R_{f_if_j} P_R + g_L \sin\zeta\, V^L_{f_if_j}
P_L\bigr) W^{\prime} f_j + \mathrm{H.c.} \,, \label{eq:lagrangian}
\eeq
where $\zeta$ is a left-right mixing angle, and $\omega$ is a
CP-violating phase that can be absorbed into $V^R$.  In this notation,
$g_{R(L)}$ are the right (left) gauge couplings, and
$V^{R,L}_{f_if_j}$ are generalized Cabibbo-Kobayashi-Maskawa (GCKM)
matrices.  In models where the $W$ and \wpr mix, the mixing angle
$\zeta$ is usually constrained to be small ($|\zeta|<$ a few $\times
10^{-5}$--$10^{-2}$ \cite{Groom:in}).  Hence, the search for purely
right- or left-handed states appears well motivated.

Many direct searches for \wpr bosons have been performed at hadron
colliders.  Most experimental analyses have considered left- or
right-handed \wpr bosons, with standard-model-like couplings, that
decay into leptonic final states.  Left-handed $W^\prime_L$ bosons, or
right-handed $W^\prime_R$ bosons in which the decay into a
right-handed neutrino $\nu_R$ is kinematically allowed, are
constrained to have masses $m_{\wprm} > 786$~GeV
\cite{Arnison:1983zy,Arnison:ut,Ansari:1987ch,Albajar:1988ka,Abe:1994zg,Abachi:1995yi,Abe:1999gz,Affolder:2001gr}.
If $m_{\nu_R} \agt m_{\wprm}$ the decay to $\nu_R$ is not allowed,
and the right-handed $W^\prime_R$ bosons are only directly constrained
by peak searches in the dijet data.  Unless the \wpr has greatly
enhanced couplings to light quarks, the dijet data are limited by QCD
backgrounds to providing a mass limit of $420$~GeV
\cite{Alitti:1990kw,Alitti:1993pn,Abe:1997hm}.

The only final state not measured by one of the experiments listed
above involves the decay of a \wpr to a single top quark.  In
Refs.~\cite{Simmons:1996ws,Tait:1997fe} it was pointed out that a
deviation of measured the single-top-quark cross section from the
standard model prediction could be evidence of a new gauge
interaction.  In the context of models with \wpr bosons, we wish to
take this one step further, and propose that the experiments search
for an explicit \wpr mass peak in the $s$-channel single-top-quark
sample.  We examine just how large the cross section into single top
quarks can be at the Fermilab Tevatron and Large Hadron Collider (LHC)
as a function of \wpr mass, and determine how enhanced or suppressed
couplings enter into the measurable cross section after cuts.  We also
determine the next-to-leading order distributions of the final-state
jets to estimate the effect on the reconstruction of the \wpr mass
peak.

A \wpr boson that propagates in the $s$-channel appears very similar
to Drell-Yan production.  We might be tempted to use the
next-to-next-to-leading order (NNLO) predictions for Drell-Yan
\cite{Hamberg:1990np} (or the updates in
Ref.~\cite{Harlander:2002wh}).  However, these calculations are not
adequate to predict the cross section for \wpr bosons which decay to
quarks.  There are two reasons for this.  The first problem is that
the final state effects are very large.  Not only are there large QCD
corrections to the quark final state, but the top-quark mass has a
large effect on the branching fraction when it is in the final state.
The second reason is that there are initial- and final-state
interference terms that arise at NNLO that are of an unknown size.
(Pure QCD processes also interfere at NNLO, e.g.\ with $u\bar d \to
u\bar d$ via the exchange of two gluons.)  Therefore, we cannot simply
take the NNLO production as calculated and append a Breit-Wigner
propagator that uses leading or next-to-leading order widths.
Instead, we perform the complete calculation at next-to-leading order.

We organize the paper as follows.  In Sec.~\ref{sec:width} we
calculate the partial widths into leptons or quarks, including the
effects of the top-quark mass, for the \wpr boson using arbitrary
vector and axial-vector couplings at next-to-leading order (NLO) in
QCD.  We demonstrate that the couplings factorize, and present some
numerical results for partial widths and branching fractions into a
top-quark final state.  In Sec.~\ref{sec:sigma} we calculate the fully
differential production and decay of a \wpr into any final state at
NLO in QCD.  We present numerical results for the single-top-quark
production cross section via $s$-channel exchange of a \wpr boson.  We
asses all theoretical uncertainties, and propose a modification of the
tolerance method for calculating parton distribution function
uncertainties.  Finally, we present the NLO jet distributions, discuss
the effect of jet definitions on these distributions, and estimate the
limits that may be placed on the \wpr mass using data from run I and
run II of the Fermilab Tevatron.

\section{Next-to-leading order width}
\label{sec:width}

We divide our evaluation of the \wpr width into three terms that
depend on the final decay products:
\beq
\Gamma_{\mathrm tot}(\wprm) = \Gamma (\wprm\to t{\bar q}^\prime)
+\Gamma (\wprm\to q{\bar q}^\prime) + \Gamma (\wprm\to l\bar\nu) \,.
\label{eq:widfact}
\eeq
We separate the partial widths containing a top quark from those
containing only massless quarks so that we may retain an explicit
top-quark mass dependence in the width.

While the large number of couplings in Eq.~(\ref{eq:lagrangian})
appears daunting, a factorization of the couplings occurs in the
partial widths of Eq.~(\ref{eq:widfact}) that greatly simplifies the
calculation.  The leading order partial widths are
\bea
\Gamma_{\mathrm LO}(\wprm\to t{\bar q}^\prime)&=&
\frac{g^2\beta^2}{16\pi m_{\wprm}} |V^\prime_{tq^\prime}|^2 (m^2_{\wprm}
+ m^2_t/2)\,, \label{eq:wlot} \\
\Gamma_{\mathrm LO}(\wprm\to q{\bar q}^\prime)&=&\frac{g^2}{16\pi}
|V^\prime_{qq^\prime}|^2 m_{\wprm} \,, \label{eq:wloq} \\
\Gamma_{\mathrm LO}(\wprm\to l\bar\nu)&=&\frac{g^2}{16\pi}
|V^\prime_{l\nu}|^2 \frac{m_{\wprm}}{3} \,, \label{eq:wlol}
\eea
where $\beta=1-m^2_t/m^2_{\wprm}$, and we assume $g^2=8 m^2_W
G_F/\sqrt{2}$, as in the standard model.  Because we have assumed there
is at most one non-zero mass in the final state, the couplings appear
only in the combination
\beq
\bigl| g V^\prime_{f_if_j}\bigr|^2 \equiv 
\bigl| g_L \sin\zeta\, V^L_{f_if_j} \bigr|^2 + \bigl| g_R
\cos\zeta\, V^R_{f_if_j} \bigr|^2 \,. \label{eq:vprime}
\eeq
Hence, the partial widths of the \wpr boson have the same form as the
standard model $W$ boson, with the effect of new couplings and GCKM
matrix elements absorbed into $V^\prime_{f_if_j}$.

The above factorization of couplings turns out to hold in the
next-to-leading order widths as well.  The calculations of the
next-to-leading order partial widths of the $W$ boson were first
performed for the massless \cite{Albert:ix}, one mass
\cite{Alvarez:1987gi}, and arbitrary mass \cite{Lahanas:1996xx} final
states many years ago.  In order to derive the factorization of
arbitrary couplings above, however, we rederive the partial widths,
but use a newer calculational method whose results are needed in the
calculation of the fully differential cross section in
Sec.~\ref{sec:sigma}.

We evaluate the next-to-leading order partial widths to quarks by
using the phase-space-slicing method with two cutoffs
\cite{Harris:2001sx}.  In this method, phase space is divided into
three regions: soft (S), hard collinear (HC), and hard non-collinear
(H$\overline{\mathrm C}$).  The first two regions are integrated over
two-body phase space in $n=4-2\epsilon$ dimensions, whereas the third
region is integrated over three-body phase space in 4 dimensions with
cuts.

The soft region of phase space is defined in the \wpr rest frame by
a condition on the energy of the emitted gluon
\begin{equation}
\label{eq:appsg}
  E_g \leq \delta_s\, \frac{m_{\wprm}}{2} \,,
\end{equation}
where $\delta_s$ is an arbitrary parameter that must cancel in the
final result.  The hard region is the complement, $E_g > \delta_s\,
m_{\wprm}/2$. The gluon can only be collinear with a light quark, and
so the collinear region is defined by comparing the invariant mass
squared of the gluon and light quark to $\delta_c m^2_{\wprm}$, i.e.\
the collinear region is $s_{qg} = 2p_q\cdot p_g < \delta_c
m^2_{\wprm}$, where $p_{q,g}$ are quark and gluon four-momenta.  The
final result must be independent of $\delta_c$.  In practice we retain
terms logarithmic in $\delta_s$ or $\delta_c$, and drop terms that are
linear in the cutoffs.  At the end, we take the couplings numerically
to zero and show the solution contains no residual $\delta_s$ or
$\delta_c$ dependence.

The two-body next-to-leading order correction to the width is given by
\beq
\delta\Gamma_{2} = \frac{\alpha_s (m_{W^\prime})}{2\pi} C_F ( M^2_S + M^2_C
+ M^2_V + \widetilde{M}^2_V ) \Gamma_{\mathrm LO} \,, \label{eq:wid2}
\eeq
where $C_F = 4/3$, and $\Gamma_{\mathrm LO}$ is listed in
Eqs.~(\ref{eq:wlot}, \ref{eq:wloq}).  If there is a top quark in the final
state, the terms are
\bea
M^2_S &=&2\ln^2(\delta_s) - 2\ln (\delta_s) \left[ 1 - \ln \left(
\frac{\mwpsq}{\mtsq} \right) \right] + \frac{\mwpsq + \mtsq}{\mwpsq -
\mtsq} \ln\! \left( \frac{\mwpsq}{\mtsq} \right) - 2 {\mathrm Li}_2(\beta)
- \frac{1}{2}\ln^2\! \left( \frac{\mwpsq}{\mtsq} \right) \,,
\label{eq:msf} \\
M^2_C &=& \frac{7}{2} - 2\zeta_2 - \ln^2(\delta_s) + 2 {\mathrm Li}_2\!
\left( \frac{\beta \delta_c}{\delta_s} \right) - \ln^2 (\beta) +2 \ln
(\delta_s) \ln (\beta) - \ln (\delta_c) \left( 2 \ln (\delta_s) +
\frac{3}{2} - 2 \ln (\beta) \right) \,, \label{eq:mcf} \\
M^2_V &=& -\frac{1}{2}\ln^2\! \left( \frac{\mwpsq}{\mtsq} \right) -
\frac{5}{2}\ln\! \left( \frac{\mwpsq}{\mtsq} \right) - \ln (1 - \lambda)
\left[ 2 + \beta + 2 \ln\! \left(
\frac{\mwpsq}{\mtsq} \right) \right] \nonumber \\
&& - 6 - \ln^2 (1 - \lambda) + 2 {\mathrm Li}_2(\lambda) - 2\zeta_2 \,, 
\label{eq:mvf} \\
\widetilde{M}^2_V &=& \frac{\beta (1-\beta)}{2 (3-\beta)}\ln(1 - \lambda) \,,
\eea
where $\zeta_2 = \pi^2/6$, $\beta=1-m^2_t/m^2_{\wprm}$, and $\lambda =
1/\beta$.  Note $(1 -\lambda) < 0$, which means $\ln (1 -\lambda) = \ln
|1 -\lambda|$, but $\ln^2 (1 -\lambda) = \ln^2 |1 -\lambda| - \pi^2$.

If both final-state quarks are massless, the formulae simplify to
\bea
M^2_S &=&4\ln^2(\delta_s) \,, \\
M^2_C &=& 7 - 4\zeta_2 - 2 \ln^2(\delta_s) + 4 {\mathrm Li}_2\! \left(
\frac{\delta_c}{\delta_s} \right) - \ln (\delta_c) [ 4 \ln (\delta_s)
+ 3 ] \,, \\
M^2_V &=& - 8 + 4 \zeta_2 \,, \\
\widetilde{M}^2_V &=& 0 \,.
\eea
The sum of the terms in the massless case is
\beq
M^2 = 2\ln^2(\delta_s) - \ln
(\delta_c) [ 4 \ln (\delta_s) + 3 ] + 4 {\mathrm Li}_2\! \left(
\frac{\delta_c}{\delta_s} \right) - 1 \,. \label{eq:mmsq}
\eeq

\newcommand{\sttp}{s^\prime_{23}}
\newcommand{\sot}{s^{}_{13}}
\newcommand{\sotp}{s^\prime_{12}}

The three-body hard non-collinear real gluon emission term is evaluated
using a Monte Carlo integration in four dimensions, subject to the
cuts listed above.  The integrand is given by
\beq
d(\delta\Gamma_{3}) = \frac{1}{64\pi^3 m_{W^\prime}}
\overline{\Sigma}|M_3|^2 dE_g dE_{q^{\prime}} \; ,
\eeq
with
\beq
\overline{\Sigma}|M_3|^2 = -2\pi\alpha_s g^2 C_F
|V^{\prime}_{tq^\prime}|^2 \Psi_\Gamma \;,
\eeq
and
\bea
\Psi_\Gamma&=& 
2\beta m_{\wprm}^2 (3-\beta)\biggl( \frac{\sotp}{\sot \sttp}
-\frac{m^2_t}{s^{\prime\,2}_{23}} \biggr) \nonumber \\
&&+ (3-\beta) \biggl(
\frac{\sot}{\sttp} + \frac{\sttp}{\sot} \biggr)
+2 (1-\beta) \;, \label{eq:appwid}
\eea
where $\beta=1-m^2_t/m^2_{\wprm}$, $\sotp = 2p_1\cdot p_2$, $\sot =
2p_1\cdot p_3$, $\sttp = 2p_2\cdot p_3$, and $p_1$, $p_2$, and $p_3$
are the four-momenta of the light quark, top quark, and gluon,
respectively.  When both quarks are massless $\beta = 1$, and
Eq.~(\ref{eq:appwid}) is very simple.

Once we determine that the factorization of couplings holds for all
terms, the partial widths reduce to the expressions given in
Ref.~\cite{Alvarez:1987gi} with the replacement of CKM matrix elements
by the GCKM matrix elements $|V^\prime|^2$ defined in
Eq.~(\ref{eq:vprime}).  The massless case is simply
\beq
\Gamma_{\mathrm NLO}(\wprm\to q{\bar q}^\prime) = \biggl( 1 +
\frac{\alpha_s(m_{\wprm})}{\pi} \biggr) \Gamma_{\mathrm LO}(\wprm\to
q{\bar q}^\prime) \,.
\eeq
The massive case should reduce to Eq.~(12) of Ref.~\cite{Alvarez:1987gi},
\bea
\Gamma_{\mathrm NLO}(\wprm\to t{\bar q}^\prime) &=&
\Gamma_{\mathrm LO}(\wprm\to t{\bar q}^\prime) + \frac{g^2}{16\pi}
\frac{\alpha_s(m_{\wprm})}{4\pi} C_F |V^\prime_{tq^\prime}|^2 m_{\wprm}
\nonumber \\
&&\hspace*{8em} \times\{ \beta(1-\beta)(\beta-4) -
\beta^2(9-5\beta)\ln(\beta) - (1-\beta)
(4+6\beta-5\beta^2)\ln(1-\beta) 
\nonumber \\
&&\hspace*{9em} + \beta^2(3-\beta) [3/2 + 4 {\mathrm
Li }_2 (1-\beta) + 2 \ln(\beta)\ln(1-\beta) ] \} \,. \label{eq:alvwid}
\eea

\subsection{Numerical results}
\label{sec:numwid}

In order to be of immediate use to experimental analyses at the
Fermilab Tevatron, we make definite predictions of the \wpr partial
and total widths.  In all numerical results we use $m_t=175\pm 5$ GeV,
$G_F=1.16639\times 10^{-5}$ GeV$^{-2}$, and $m_W=80.4$ GeV.  We use a
two-loop running of $\alpha_s$ as defined in the CTEQ5M1 parton
distribution functions \cite{Lai:1999wy}.  We assume that
$V^\prime_{l\nu}$ is the identity matrix, and for
$V^\prime_{qq^\prime}$ we use the average Cabibbo-Kobayashi-Maskawa
(CKM) matrix \cite{Groom:in} with the exception that we assume
$|V^\prime_{tb}|=1$,
\beq
V^\prime_{\mathrm CKM} = \left(
\begin{array}{llc}
0.9751 & 0.2215 & 0.0035 \\
0.2210 & 0.9743 & 0.0410 \\
0.0090 & 0.0400 & 1.0000
\end{array}
\right) \,. \label{eq:ckm}
\eeq

\begin{figure}[tbh]
\epsfxsize=3.375in
\centerline{\epsfbox{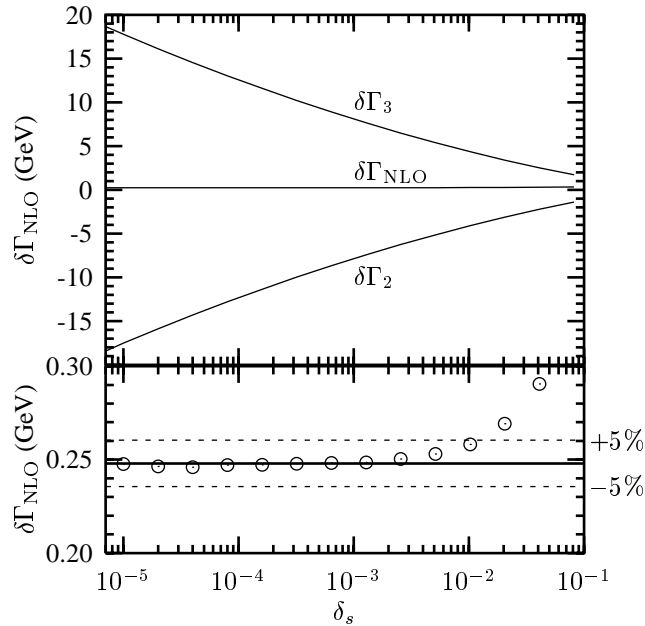}} % 243pt = 3.375in actual size
\caption{Dependence of the NLO correction to the \wpr width on
$\delta_s = 300\times \delta_c$, for $m_{W^{\prime}}=500$ GeV.
$\delta\Gamma_{\mathrm NLO} = \delta\Gamma_2 + \delta\Gamma_3$, where
$\delta\Gamma_3$ refers to the H$\overline{\mathrm{C}}$ component, and
$\delta\Gamma_2$ to the sum of virtual, soft and hard-collinear
contributions.  The bottom region shows an enlargement of the
numerical prediction (circles) compared to the analytic prediction
(solid line) in Eq.~(\protect\ref{eq:alvwid}).\label{fig:wpss}}
\end{figure}

Before we discuss our results, we must show that our final result does
not depend on the cutoffs we choose.  To demonstrate this, we fix the
ratio of $\delta_s/\delta_c = 300$, and plot in Fig.~\ref{fig:wpss}
the NLO correction to the width of a \wpr of mass 500~GeV.  In the
upper half of the figure we see the logarithmic dependence in
$\delta_s$ and $\delta_c$ of the two- and three-body terms,
$\delta\Gamma_2$ and $\delta\Gamma_3$.  In the bottom half of the
figure we focus on the sum of the correction terms
$\delta\Gamma_{\mathrm NLO}$ for various values of $\delta_s$
(circles), and compare the result to the analytic prediction (solid
line) of Eq.~(\ref{eq:alvwid}).  Once $\delta_s \alt 10^{-3}$
($\delta_c \alt 1/3\times 10^{-5}$), the result is stable to much less
than $1\%$ of the known NLO correction.
In practice there is a tradeoff between numerical accuracy in
canceling the logarithmic divergences of $\delta_s$ and $\delta_c$ in
the Monte Carlo, and the residual power suppressed dependence on these
parameters.  Hence, we always choose values of $\delta_s$ and
$\delta_c$ to ensure that the residual effects are smaller than the
desired Monte Carlo statistical error.

In Table~\ref{tab:pwid} we list the LO and NLO partial widths for the
decay of the \wpr into light quarks, or $t\bar b$.  Since we assume
the standard model-like couplings of Eq.~(\ref{eq:ckm}), the decays
into $t\bar d$ and $t\bar s$ are strongly suppressed.  However, we
remind the reader that any model may be restored via the use of
Eq.~(\ref{eq:vprime}).  We present the results for \wpr masses between
200 GeV and 1000 GeV, in order to cover the possible reach of
experiments at the Tevatron.

\begin{table}[htb]
\caption{LO and NLO partial widths for \wpr decays into light quarks,
or $t\bar b$.\label{tab:pwid}}
\begin{center}
\begin{tabular}{ddddd}
Mass (GeV)&$\Gamma_{\mathrm LO} (W^\prime\to q{\bar q}^\prime)$ (GeV)&
$\Gamma_{\mathrm NLO} (W^\prime\to q{\bar q}^\prime)$ (GeV)&
$\Gamma_{\mathrm LO} (W^\prime\to t\bar b)$ (GeV)&
$\Gamma_{\mathrm NLO} (W^\prime\to t\bar b)$ (GeV)\\\hline \\
200. & 1.697 & 1.754 & 0.129 & 0.166 \\ \\
225. & 1.909 & 1.973 & 0.388 & 0.473 \\ \\
250. & 2.121 & 2.191 & 0.687 & 0.811 \\ \\
275. & 2.333 & 2.409 & 0.993 & 1.146 \\ \\
300. & 2.546 & 2.628 & 1.296 & 1.471 \\ \\
350. & 2.970 & 3.064 & 1.879 & 2.084 \\ \\
400. & 3.394 & 3.500 & 2.432 & 2.655 \\ \\
450. & 3.818 & 3.936 & 2.959 & 3.196 \\ \\
500. & 4.243 & 4.372 & 3.467 & 3.715 \\ \\
550. & 4.667 & 4.807 & 3.961 & 4.218 \\ \\
600. & 5.091 & 5.243 & 4.443 & 4.710 \\ \\
650. & 5.515 & 5.678 & 4.917 & 5.192 \\ \\
700. & 5.940 & 6.114 & 5.384 & 5.667 \\ \\
750. & 6.364 & 6.549 & 5.845 & 6.136 \\ \\
800. & 6.788 & 6.984 & 6.301 & 6.600 \\ \\
850. & 7.212 & 7.420 & 6.754 & 7.061 \\ \\
900. & 7.637 & 7.855 & 7.204 & 7.518 \\ \\
950. & 8.061 & 8.290 & 7.651 & 7.972 \\ \\
1000.& 8.485 & 8.725 & 8.096 & 8.425
\end{tabular}
\end{center}
\end{table}

Unless the couplings are suppressed, a left-handed \wpr will decay
into either quarks or leptons.  A right-handed $\wprm$, however, will
decay to leptons only if the mass of the right-handed neutrino
($m_{\nu_R}$) is less than $m_{\wprm}$, or if there is large
left-right mixing in the neutrino sector.  For completeness, we
consider both kinematic cases.  In Tables~\ref{tab:widq} and
\ref{tab:widql} we list the LO and NLO \wpr total widths and branching
fractions into a $t\bar b$ final state.  We assume a top-quark mass of
175 GeV.  The largest uncertainty in the predictions of both the total
width and the branching fractions is a result of the uncertainty in
the top-quark mass.  Hence, we also show the increase (upper error),
or decrease (lower error), in the quantities if the top-quark mass is
170 GeV, or 180 GeV, respectively.  In Table~\ref{tab:widq} we include
only the decays into quark final states.  In Table~\ref{tab:widql} we
allow for the decay into quarks or leptons as in the standard model.
The total widths in these Tables are used in Sec.~\ref{sec:sigma} in
the NLO calculation of the \wpr contribution to the single-top-quark
cross section.

\begin{table}[htb]
\caption{Total width of the $W^\prime$, and its branching ratio into
$t\bar b$, at LO and NLO in QCD, when the decay to leptons is not
allowed.  Errors are due to the top-quark mass uncertainty, $m_t=175
\mp 5$~GeV.\label{tab:widq}}
\begin{center}
\begin{tabular}{dr@{ }lr@{ }lr@{ }lr@{ }l}
Mass (GeV)&\multicolumn{2}{c}{$\Gamma_{\mathrm LO}$ (GeV)}&
\multicolumn{2}{c}{BR$_{\mathrm LO}$($W^\prime\to t\bar b$)}&
\multicolumn{2}{c}{$\Gamma_{\mathrm NLO}$ (GeV)}&
\multicolumn{2}{c}{BR$_{\mathrm NLO}$($W^\prime\to t\bar b$)}\\\hline \\
200. & 3.523&$^{+0.049}_{-0.043}$ & 0.0366&$^{+0.0132}_{-0.0119}$ & 3.675&$^{+0.060}_{-0.053}$ & 0.0452&$^{+0.0152}_{-0.0140}$ \\ \\
225. & 4.206&$^{+0.064}_{-0.061}$ & 0.0923&$^{+0.0136}_{-0.0135}$ & 4.419&$^{+0.073}_{-0.071}$ & 0.1071&$^{+0.0144}_{-0.0146}$ \\ \\
250. & 4.930&$^{+0.068}_{-0.067}$ & 0.1394&$^{+0.0117}_{-0.0119}$ & 5.193&$^{+0.074}_{-0.074}$ & 0.1561&$^{+0.0119}_{-0.0122}$ \\ \\
275. & 5.660&$^{+0.067}_{-0.068}$ & 0.1755&$^{+0.0097}_{-0.0100}$ & 5.965&$^{+0.071}_{-0.072}$ & 0.1922&$^{+0.0095}_{-0.0099}$ \\ \\
300. & 6.388&$^{+0.065}_{-0.066}$ & 0.2030&$^{+0.0081}_{-0.0083}$ & 6.726&$^{+0.067}_{-0.068}$ & 0.2187&$^{+0.0077}_{-0.0080}$ \\ \\
350. & 7.819&$^{+0.059}_{-0.060}$ & 0.2404&$^{+0.0057}_{-0.0059}$ & 8.211&$^{+0.058}_{-0.060}$ & 0.2537&$^{+0.0053}_{-0.0055}$ \\ \\
400. & 9.220&$^{+0.053}_{-0.054}$ & 0.2637&$^{+0.0042}_{-0.0044}$ & 9.654&$^{+0.051}_{-0.052}$ & 0.2750&$^{+0.0038}_{-0.0040}$ \\ \\
450. & 10.595&$^{+0.048}_{-0.049}$ & 0.2792&$^{+0.0032}_{-0.0033}$ & 11.068&$^{+0.045}_{-0.046}$ & 0.2888&$^{+0.0029}_{-0.0030}$ \\ \\
500. & 11.952&$^{+0.043}_{-0.044}$ & 0.2901&$^{+0.0026}_{-0.0027}$ & 12.458&$^{+0.040}_{-0.041}$ & 0.2982&$^{+0.0022}_{-0.0023}$ \\ \\
550. & 13.294&$^{+0.040}_{-0.041}$ & 0.2979&$^{+0.0021}_{-0.0022}$ & 13.832&$^{+0.036}_{-0.037}$ & 0.3049&$^{+0.0018}_{-0.0019}$ \\ \\
600. & 14.625&$^{+0.036}_{-0.037}$ & 0.3038&$^{+0.0017}_{-0.0018}$ & 15.195&$^{+0.032}_{-0.034}$ & 0.3099&$^{+0.0015}_{-0.0015}$ \\ \\
650. & 15.947&$^{+0.034}_{-0.035}$ & 0.3083&$^{+0.0015}_{-0.0015}$ & 16.548&$^{+0.030}_{-0.031}$ & 0.3137&$^{+0.0012}_{-0.0013}$ \\ \\
700. & 17.263&$^{+0.031}_{-0.032}$ & 0.3119&$^{+0.0012}_{-0.0013}$ & 17.894&$^{+0.027}_{-0.028}$ & 0.3167&$^{+0.0010}_{-0.0011}$ \\ \\
750. & 18.572&$^{+0.029}_{-0.030}$ & 0.3147&$^{+0.0011}_{-0.0011}$ & 19.234&$^{+0.025}_{-0.026}$ & 0.3190&$^{+0.0009}_{-0.0009}$ \\ \\
800. & 19.878&$^{+0.027}_{-0.028}$ & 0.3170&$^{+0.0009}_{-0.0010}$ & 20.569&$^{+0.023}_{-0.024}$ & 0.3209&$^{+0.0008}_{-0.0008}$ \\ \\
850. & 21.179&$^{+0.026}_{-0.026}$ & 0.3189&$^{+0.0008}_{-0.0009}$ & 21.900&$^{+0.021}_{-0.023}$ & 0.3224&$^{+0.0007}_{-0.0007}$ \\ \\
900. & 22.477&$^{+0.024}_{-0.025}$ & 0.3205&$^{+0.0007}_{-0.0008}$ & 23.227&$^{+0.021}_{-0.021}$ & 0.3237&$^{+0.0006}_{-0.0006}$ \\ \\
950. & 23.773&$^{+0.023}_{-0.024}$ & 0.3218&$^{+0.0007}_{-0.0007}$ & 24.552&$^{+0.020}_{-0.019}$ & 0.3247&$^{+0.0005}_{-0.0005}$ \\ \\
1000.& 25.066&$^{+0.022}_{-0.023}$ & 0.3230&$^{+0.0006}_{-0.0006}$ & 25.875&$^{+0.018}_{-0.018}$ & 0.3256&$^{+0.0005}_{-0.0005}$
\end{tabular}
\end{center}
\end{table}

\begin{table}[htb]
\caption{Total width of the $W^\prime$, and its branching ratio into
$t\bar b$, at LO and NLO in QCD, when decays to quarks or leptons are
both included.  Errors are due to the top-quark mass uncertainty,
$m_t=175 \mp 5$~GeV.\label{tab:widql}}
\begin{center}
\begin{tabular}{dr@{ }lr@{ }lr@{ }lr@{ }l}
Mass (GeV)&\multicolumn{2}{c}{$\Gamma_{\mathrm LO}$ (GeV)}&
\multicolumn{2}{c}{BR$_{\mathrm LO}$($W^\prime\to t\bar b$)}&
\multicolumn{2}{c}{$\Gamma_{\mathrm NLO}$ (GeV)}&
\multicolumn{2}{c}{BR$_{\mathrm NLO}$($W^\prime\to t\bar b$)}\\\hline \\
200. & 5.220&$^{+0.049}_{-0.043}$ & 0.0247&$^{+0.0091}_{-0.0081}$ & 5.372&$^{+0.060}_{-0.053}$ & 0.0309&$^{+0.0106}_{-0.0097}$ \\ \\
225. & 6.116&$^{+0.064}_{-0.061}$ & 0.0635&$^{+0.0097}_{-0.0095}$ & 6.328&$^{+0.073}_{-0.071}$ & 0.0748&$^{+0.0105}_{-0.0105}$ \\ \\
250. & 7.051&$^{+0.068}_{-0.067}$ & 0.0974&$^{+0.0086}_{-0.0087}$ & 7.314&$^{+0.074}_{-0.074}$ & 0.1109&$^{+0.0089}_{-0.0091}$ \\ \\
275. & 7.994&$^{+0.067}_{-0.068}$ & 0.1243&$^{+0.0073}_{-0.0075}$ & 8.298&$^{+0.071}_{-0.072}$ & 0.1381&$^{+0.0073}_{-0.0076}$ \\ \\
300. & 8.933&$^{+0.065}_{-0.066}$ & 0.1451&$^{+0.0062}_{-0.0064}$ & 9.272&$^{+0.067}_{-0.068}$ & 0.1587&$^{+0.0060}_{-0.0063}$ \\ \\
350. & 10.789&$^{+0.059}_{-0.060}$ & 0.1742&$^{+0.0045}_{-0.0046}$ & 11.181&$^{+0.058}_{-0.060}$ & 0.1863&$^{+0.0042}_{-0.0044}$ \\ \\
400. & 12.614&$^{+0.053}_{-0.054}$ & 0.1928&$^{+0.0034}_{-0.0035}$ & 13.049&$^{+0.051}_{-0.052}$ & 0.2035&$^{+0.0031}_{-0.0032}$ \\ \\
450. & 14.414&$^{+0.048}_{-0.049}$ & 0.2053&$^{+0.0026}_{-0.0027}$ & 14.886&$^{+0.045}_{-0.046}$ & 0.2147&$^{+0.0023}_{-0.0024}$ \\ \\
500. & 16.195&$^{+0.043}_{-0.044}$ & 0.2141&$^{+0.0021}_{-0.0022}$ & 16.701&$^{+0.040}_{-0.041}$ & 0.2224&$^{+0.0019}_{-0.0019}$ \\ \\
550. & 17.961&$^{+0.040}_{-0.041}$ & 0.2205&$^{+0.0017}_{-0.0018}$ & 18.499&$^{+0.036}_{-0.037}$ & 0.2280&$^{+0.0015}_{-0.0015}$ \\ \\
600. & 19.716&$^{+0.036}_{-0.037}$ & 0.2253&$^{+0.0014}_{-0.0015}$ & 20.287&$^{+0.032}_{-0.034}$ & 0.2322&$^{+0.0012}_{-0.0013}$ \\ \\
650. & 21.463&$^{+0.034}_{-0.035}$ & 0.2291&$^{+0.0012}_{-0.0012}$ & 22.064&$^{+0.030}_{-0.031}$ & 0.2353&$^{+0.0010}_{-0.0011}$ \\ \\
700. & 23.202&$^{+0.031}_{-0.032}$ & 0.2320&$^{+0.0010}_{-0.0011}$ & 23.834&$^{+0.027}_{-0.028}$ & 0.2378&$^{+0.0009}_{-0.0009}$ \\ \\
750. & 24.936&$^{+0.029}_{-0.030}$ & 0.2344&$^{+0.0009}_{-0.0009}$ & 25.598&$^{+0.025}_{-0.026}$ & 0.2397&$^{+0.0007}_{-0.0008}$ \\ \\
800. & 26.666&$^{+0.027}_{-0.028}$ & 0.2363&$^{+0.0008}_{-0.0008}$ & 27.357&$^{+0.023}_{-0.024}$ & 0.2413&$^{+0.0006}_{-0.0007}$ \\ \\
850. & 28.391&$^{+0.026}_{-0.026}$ & 0.2379&$^{+0.0007}_{-0.0007}$ & 29.113&$^{+0.021}_{-0.023}$ & 0.2425&$^{+0.0006}_{-0.0006}$ \\ \\
900. & 30.114&$^{+0.024}_{-0.025}$ & 0.2392&$^{+0.0006}_{-0.0006}$ & 30.864&$^{+0.021}_{-0.021}$ & 0.2436&$^{+0.0005}_{-0.0005}$ \\ \\
950. & 31.834&$^{+0.023}_{-0.024}$ & 0.2403&$^{+0.0006}_{-0.0006}$ & 32.613&$^{+0.020}_{-0.019}$ & 0.2444&$^{+0.0005}_{-0.0004}$ \\ \\
1000.& 33.551&$^{+0.022}_{-0.023}$ & 0.2413&$^{+0.0005}_{-0.0005}$ & 34.360&$^{+0.018}_{-0.018}$ & 0.2452&$^{+0.0004}_{-0.0004}$
\end{tabular}
\end{center}
\end{table}

It is not surprising that top-quark threshold effects cause a large
uncertainty in the branching fraction for \wpr masses less than 300
GeV.  This also appears in the large increase in branching fraction at
NLO over the LO branching fraction.  With 33\% changes in both the
mass variation and NLO correction, perturbation theory is somewhat
suspect if $m_{\wprm} \sim 200$ GeV.  However, the effect is less than
10\% by 225 GeV, and rapidly vanishes as $m^2_{\wprm} \gg m^2_t$.  The
branching fraction into a top-quark final state is nearly saturated at
$1/3$ ($1/4$) when $m_{\wprm} \sim 500$~GeV for the quark-only
(quark-plus-lepton) model of decays.  Hence, a large fraction of \wpr
events should produce a top-quark in the final state.  In
Sec.~\ref{sec:tottop} we see this effect on the single-top-quark cross
section.

\section{Next-to-leading order cross section}
\label{sec:sigma}

The analytic form of the differential production and decay of a \wpr
at next-to-leading order is very similar to that of $s$-channel
single-top-quark production.  The complete calculation of differential
single-top-quark production using the phase space slicing method
appears in Ref.~\cite{Harris:2002md}.  We generalize the calculation
in that paper to the production of one massive particle through a \wpr
boson (such that single-top-quark production is a special case).  We
follow closely the notation of Ref.~\cite{Harris:2002md}, but retain
arbitrary couplings in the vertices.  We note that the $t$-channel
exchange of a virtual \wpr is suppressed by at least
$1/m^4_{W^\prime}$, and hence do not consider it here.  However, the
analytic expressions and factorizations that follow are valid for this
channel as well.  The analytic formulae for the $t$-channel exchange
may be obtained by simple crossing.

At leading order, the production of a heavy quark (lepton) may be
written schematically as
\begin{equation}
u(p_1)\bar d(p_2) \to \wprmp \to \bar b(p_3) t(p_4) [\bar l(p_3)
\nu(p_4)] \,,
\end{equation}
where $p_1+p_2=p_3+p_4$, $u\bar d$ represents all possible parton
fluxes, and the heavy particle in the final state ($t$ or $\nu_R$) has
four-momentum $p_4$ and mass $m$.

The leading-order fully differential spin-averaged partonic cross
section can be written as
\begin{equation}
d\sigma_{0} = \frac{1}{2s} \overline{\sum}
\left|{\cal M}_{0}\right|^2 d {\mathrm{PS}}_2 \,,
\end{equation}
where $s=(p_1+p_2)^2$ is the partonic center-of-momentum energy
squared, and the two body phase space is given by
\begin{equation}
d{\mathrm{PS}}_2 = \frac{1}{(2\pi)^2} \frac{d^3 p_3}{2 E_3} \frac{d^3
   p_4}{2 E_4} \delta^{(4)}(q-p_3-p_4)\, .
\end{equation}

The matrix element squared, summed (averaged) over final- (initial-)
state spin and color states, for production of one massive particle is
\begin{equation}
\overline{\sum} \left|{\cal M}_{0}\right|^2 = \frac{2N_c}{3} \frac{R_t
t\bigl( t-m^2\bigr) + R_u u\bigl( u-m^2\bigr)}{\bigl( s-\mwpsq
\bigr)^2 +\mwpsq \Gamma_{W^\prime}^2} \,,
\end{equation}
where $t=(p_1-p_3)^2$, $u=(p_2-p_3)^2$. The number of colors $N_c=3$,
and $m = m_t$, for a final state with a top quark.  For the production
of a massive neutrino $m = m_{\nu_R}$ and $N_c=1$.  If the final state
is massless, simply set $m=0$ and use $N_c=3$ (1) for quarks
(leptons).  The functions $R_t$ and $R_u$ are given by
\bea
R_t&=&\bigl( |V_i|^2+|A_i|^2 \bigr) \bigl( |V_f|^2+|A_f|^2 \bigr) +
4 {\mathrm Re } (V_i{A_i}^{\!\ast}) {\mathrm Re } (V_f\!{A_f}^{\!\ast}) \,,
\label{eq:rt} \\
R_u&=&\bigl( |V_i|^2+|A_i|^2 \bigr) \bigl( |V_f|^2+|A_f|^2 \bigr) -
4 {\mathrm Re } (V_i{A_i}^{\!\ast}) {\mathrm Re } (V_f\!{A_f}^{\!\ast}) \,,
%R_u&=&\bigl( V_i^2+A_i^2 \bigr) \bigl( V_f^2+A_f^2 \bigr) - 4V_iV_fA_iA_f
\label{eq:ru}
\eea
where $V_{i,f}$ and $A_{i,f}$ are the vector and axial-vector
couplings for the initial- and final-state vertices, respectively.  In
the notation of Eq.~(\ref{eq:lagrangian}),
\bea
|V|^2+|A|^2 &=& \frac{\bigl| g_R \cos\zeta\, V^R_{f_if_j} \bigr|^2
+\bigl| g_L \sin\zeta\, V^L_{f_if_j} \bigr|^2}{4} \,, \\
2 {\mathrm Re } (V\! A^{\ast}) &=& \frac{\bigl| g_R \cos\zeta\, V^R_{f_if_j}
\bigr|^2 -\bigl| g_L \sin\zeta\, V^L_{f_if_j} \bigr|^2}{4} \,.
\eea
For right- or left-handed \wpr bosons $|V|=|A|$, $R_u = 0$, $R_t$
reduces to $R_t = \bigl( g^4 |V^\prime_i|^2 |V^\prime_f|^2 \bigr) /8$,
and we recover the standard model $s$-channel single-top-quark cross
section published in Refs.~\cite{Harris:2002md,Harris:2000sv},
including the Breit-Wigner term, up to the GCKM couplings.

We calculate the fully differential next-to-leading order cross
section with the same method used in Sec.~\ref{sec:width}.
Again, the phase space is divided into a hard and soft region, where
the soft region is now defined in the partonic center-of-momentum
frame by a condition of the emitted gluon energy:
\begin{equation}
  E_g \leq \delta_s\, \frac{\sqrt{s_{12}}}{2} \,,
\end{equation}
where $s_{ij}=(p_i+p_j)^2$.  The hard region is the complement, $E_g >
\delta_s\, \sqrt{s_{12}}/2$.  A region is collinear if an invariant
in a propagator is less than $\delta_c s_{12}$, i.e.\ the region is
collinear if in a denominator we obtain $s_{ij} = (p_i+p_j)^2 <
\delta_c s_{12}$, or $t_{ij} = (p_i-p_j)^2 < \delta_c s_{12}$.

At next-to-leading order, color conservation forbids the exchange of a
single gluon between the initial and final states.  This is convenient
when organizing the solution, because the leptonic cross section only
has QCD corrections in the initial state.  Hence, we list separately
the initial and final state corrections.

The two-body NLO correction to the partonic cross section contains
\begin{equation}
d\sigma_2 = \frac{\alpha_s}{2\pi} C_F [(M_S^2 + M_C^2 + M_V^2)d\sigma_0
+d\tilde{\sigma}_V] \,, \label{eq:dsig2}
\end{equation}
where $C_F=4/3$, and $d\tilde{\sigma}_V$ is the part of the correction
to the vertex containing the top quark that is not proportional to the
LO cross section.  If the final state does not contain a top quark
then $d\tilde{\sigma}_V=0$; otherwise,
\begin{equation}
d\tilde{\sigma}_V = (R_t + R_u) \frac{t u}{s^2}\frac{m_t^2 \ln(1-
\lambda)}{\bigl( s-\mwpsq\bigr)^2 +\mwpsq \Gamma_{W^\prime}^2} 
d{\mathrm{PS}}_2 \,.
\end{equation}

The initial-state corrections appearing in Eq.~(\ref{eq:dsig2}) are
\bea
M^2_{S i} &=& 4\ln^2(\delta_s) \,, \\
M^2_{C i} &=& [4\ln(\delta_s) + 3] \ln\!\left( \frac{s}{\mu^2_F}\right)
\,, \\
M^2_{V i} &=& -8+4\zeta_2 \,,
\eea
where $\zeta_2 = \pi^2/6$, and $\mu_F$ is the factorization scale.
If there is a top quark in the final state, $M^2_{S f}$, $M^2_{C f}$,
and $M^2_{V f}$ are the same as those listed in
Eqs.~(\ref{eq:msf}--\ref{eq:mvf}), with the replacements $\mwpsq \to s$,
$\beta \to 1-m_t^2/s$, and $\lambda = 1/\beta$.  If both quarks in the
final state are massless, then $M^2_f$ is given by Eq.~(\ref{eq:mmsq}).

The complete two-body cross section is given by
\begin{equation}
\sigma_2 = \sum_{a,b} \int dx_1 dx_2 \{ f_a^{H_1}(x_1,\mu_F)
f_b^{H_2}(x_2,\mu_F) d\sigma_2^{a b} + [ \tilde{f}_a^{H_1}(x_1,\mu_F)
f_b^{H_2}(x_2,\mu_F) + f_a^{H_1}(x_1,\mu_F)
\tilde{f}_b^{H_2}(x_2,\mu_F) ] d\sigma_0^{a b} \} \,,
\end{equation}
where $a$, $b$ sum over all quark-antiquark luminosities, $\mu_F$ is
the factorization scale, $H_{1,2}$ are the initial-state hadrons, and
the $f(x,\mu)$ are NLO parton distribution functions.  The $\tilde{f}$
functions are introduced to compensate for a difference between the
limits of integration used in the phase space slicing calculation of
the initial-state collinear singularities and the modified minimal
subtraction $\overline{\mathrm MS}$ scheme used in the NLO parton
distribution functions.  The $\tilde{f}$ functions are given in
Ref.~\cite{Harris:2001sx}.

%%%%%%%%%%%%%%%%%%%% New commands for 3-body Eqs. %%%%%%%%%%%%%%%%%%%
\renewcommand{\sot}{s^{}_{12}}
\newcommand{\stfp}{s^{\prime}_{34}}
\newcommand{\stf}{s^{}_{35}}
\newcommand{\sffp}{s^{\prime}_{45}}
\newcommand{\tot}{t^{}_{13}}
\newcommand{\ttt}{t^{}_{23}}
\newcommand{\tofp}{t^{\prime}_{14}}
\newcommand{\ttfp}{t^{\prime}_{24}}
\newcommand{\tof}{t^{}_{15}}
\newcommand{\ttf}{t^{}_{25}}
%%%%%%%%%%%%%%%%%%%% New commands for 3-body Eqs. %%%%%%%%%%%%%%%%%%%

The three-body hard non-collinear terms are evaluated using a Monte
Carlo integration in four dimensions, subject to the cuts listed
above.  The cross section is given by
\begin{eqnarray}
\sigma_3 &=& - \sum_{a,b} \int dx_1 dx_2 \frac{4\pi\alpha_s}{s}
\int_{\mathrm H\overline{C}} \{ f^{H_1}_{q_a}(x_1,\mu_F)
f^{H_2}_{q_b}(x_2,\mu_F) \Psi_{q\bar q} \nonumber \\
&&\hspace*{12em} + [ f^{H_1}_{q_a}(x_1,\mu_F) f^{H_2}_{g}(x_2,\mu_F) +
f^{H_1}_{g}(x_1,\mu_F) f^{H_2}_{q_b}(x_2,\mu_F) ] \Psi_{qg} \}
d{\mathrm{PS}}_3 \,.
\end{eqnarray}
If we label the momenta in the three-body processes
\bea
\Psi_{q\bar q}: u(p_1) \bar d(p_2) &\to & \bar b(p_3) t(p_4) g(p_5) \,, \\
\Psi_{qg}: u(p_1) g(p_2) &\to & \bar b(p_3) t(p_4) d(p_5) \,,
\eea
then functions $\Psi_{q\bar q}$ and $\Psi_{qg}$ are given by
\bea
\Psi_{q\bar q}&=& \frac{2 C_F}{(s^{}_{34} - \mwpsq)^2 +\mwpsq
\Gamma_{W^\prime}^2} \left[ R_t \left( -\frac{\ttfp\left( \tot+\stf\right)
- \tot\tofp}{\tof} -\frac{\tot\left( \ttfp+\sffp\right) -
\ttt\ttfp}{\ttf}\right. \right.  \nonumber \\
&&\hspace*{14em}
\left. \left. -\frac{\sot\left[ \tot\left( 2\ttfp+\sffp\right) +
\stf\ttfp\right] }{\tof\ttf} \right) + R_u
\left( p_1 \rightleftharpoons p_2 \right) \right] \nonumber \\
&&+ \frac{2 C_F}{(\sot - \mwpsq)^2 +\mwpsq \Gamma_{W^\prime}^2} \left[
R_t \left( -\frac{\ttfp\left( \tot+\tof\right) -\tot\ttt}{\stf}
-\frac{\tot\left( \ttfp+\ttf\right) \left( 1 - 2 m_t^2/\sffp
\right) - \tofp\ttfp}{\sffp}\right. \right. \nonumber \\
&&\hspace*{15em}
\left. \left. -\frac{\stfp\left[ \tot\left( 2\ttfp+\ttf\right) +
\tof\ttfp\right] }{\stf\sffp} \right) + R_u
\left( p_1 \rightleftharpoons p_2 \right) \right] \;, \label{eq:psiqq} \\
\Psi_{qg}&=& \frac{1}{(s^{}_{34} - \mwpsq)^2 +\mwpsq \Gamma_{W^\prime}^2}
\left[ R_t \left( \frac{\sffp\left( \tot+\ttt\right) -\tot\tofp}{\sot}
+\frac{\tot\left( \sffp+\ttfp\right) - \stf\sffp}{\ttf} \right. \right.
\nonumber \\
&&\hspace*{14em} \left. \left. +\frac{\tof\left[ \sffp\left( 2\tot
+\ttt\right) +\tot\ttfp\right]}{\sot\ttf} \right) \right. \nonumber \\
&&\hspace*{12em} + R_u
\left( \frac{\stf\left( \tofp+\ttfp\right) -\tot\tofp}{\sot}
+\frac{\tofp\left( \stf+\ttt\right) - \stf\sffp}{\ttf} \right.
\nonumber \\
&&\hspace*{15em} \left. \left. +\frac{\tof\left[ \stf\left( 2\tofp
+\ttfp\right) +\tofp\ttt\right]}{\sot\ttf} \right) \right] \;,
\label{eq:psiqg}
\eea
where $s^{\prime}_{ij}= s_{ij}-m_t^2$, $t^{\prime}_{ij}=t_{ij}-m_t^2$,
and all other terms may be obtained by crossing.  Final-state QCD
corrections are limited to the second term of $\Psi_{q\bar q}$, and
thus this term does not appear in corrections to the leptonic final
state.  If the final state is massless, then the solution is recovered
by setting $m_t=0$, and noting that all primed invariants are equal to
their unprimed counterparts.

\subsection{Single-top-quark production via $W^\prime_{R,\,L}$}
\label{sec:tottop}

\begin{figure}[tbh]
\epsfxsize=3.4in
\centerline{\epsfbox{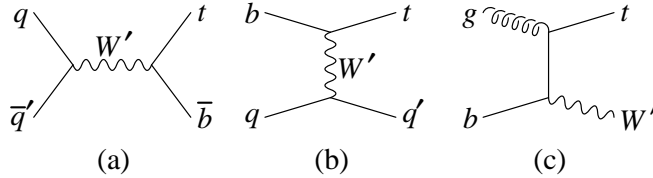}} % 268pt actual size
\caption{Representative Feynman diagrams for single-top-quark
production involving a \wpr boson: (a) $s$-channel production, (b)
$t$-channel production, and (c) $\wprm \!-t$ associated
production.\label{fig:feyn}}
\end{figure}

In Sec.~\ref{sec:numwid} we saw that \wpr bosons tend to have a large
branching fraction into top quarks.  This observation leads us to
consider the effect of such a \wpr on the size of the single-top-quark
cross section at hadron colliders.  In particular, we show the cross
sections are large enough to improve the mass limits on \wpr bosons
using data from run I of the Fermilab Tevatron.  The \wpr boson affects
the single-top-quark cross section through the three channels shown in
Fig.~\ref{fig:feyn}.  We concentrate on the $s$-channel production,
because of the enhancement from the \wpr resonance.  We do not present
numerical results for the $t$-channel or associated production of a
\wpr boson, because the cross sections for these channels are
negligible for the masses we consider.

In order to make definitive predictions, we stick to the assumption
that the \wpr has purely right- or left-handed interactions.  We
choose the standard model CKM matrix as in Eq.~(\ref{eq:ckm}), with
the understanding that the final results may be scaled using the
factorization of couplings in the last section.  As in
Sec.~\ref{sec:numwid}, we choose $m_t=175\pm 5$~GeV,
$G_F=1.16639\times 10^{-5}$ GeV$^{-2}$, and $m_W=80.4$ GeV.  For
leading-order cross sections we use CTEQ5L \cite{Lai:1999wy}
leading-order parton distribution functions (PDFs).  We use a two-loop
running of $\alpha_s$, and CTEQ5M1 PDFs for NLO cross sections.  We
use the LO and NLO widths calculated in Sec.~\ref{sec:numwid} in the
LO and NLO cross sections, respectively, so that all terms are
calculated at the same order and with the same method.  We set both
factorization and renormalization scales equal to $m_{\wprm}$, since
most of the cross section will be produced near resonance.

We present the leading- and next-to-leading order cross sections for
single-top-quark production, via $W^\prime_{R,\,L}$ bosons of various
masses, at run I of the Tevatron (a $p\bar p$ collider with
$\sqrt{S}=1.8$~TeV) in Tables~\ref{tab:sigq} and \ref{tab:sigql}.  In
Table~\ref{tab:sigq} we perform the calculation under the assumption
that $\nu_R$ is too heavy for the \wpr to decay to leptons, and thus
use the widths of Table~\ref{tab:widq}.  In Table~\ref{tab:sigql} we
assume all decays are allowed, and use the widths of
Table~\ref{tab:widql}.  The cross sections for just top-quark or
antitop-quark production are one-half of the listed results.  The
corresponding single-top-quark cross sections at a run II of the
Tevatron (a $p\bar p$ collider with $\sqrt{S}=1.96$~TeV) appear in
Tables~\ref{tab:sigqt} and \ref{tab:sigqlt}.

In Tables~\ref{tab:sigq}--\ref{tab:sigqlt} we show the uncertainties
in the cross sections due to the variation of the scale between
$m_{\wprm}/2$ and $2m_{\wprm}$, the variation of the top-quark mass
over $m_t=175 \pm 5$ GeV including the effects of the change in width
listed in Tables~\ref{tab:widq} and \ref{tab:widql}, and uncertainties
from the parton distribution functions (using a new method described
in Sec.~\ref{sec:pdfu}).  Additional uncertainties come from the use
of a vanishing bottom-quark mass, which overestimates the cross
section by $+1.5\%$ at 200 GeV, but only $+0.4\%$ by 225 GeV, and is a
negligible effect at higher \wpr masses.  Extrapolating from the
results of Ref.~\cite{Smith:1996ij}, we estimate that Yukawa
corrections cause a shift of less than $-1\%$ in the cross section.
The uncertainty from Monte Carlo statistics is $0.03\%$.  There is a
$\pm 4\%$ uncertainty in the NLO correction term from $\alpha_s =
0.118\pm 0.005$.  All uncertainties listed above are added in
quadrature, and are presented for easy reference in the last column of
each section in Tables~\ref{tab:sigq}--\ref{tab:sigqlt} as a
percentage error.  If we had used the narrow width approximation, and
multiplied the NLO \wpr cross section times NLO branching fraction to
$t\bar b$ instead of evaluating the full matrix element, there would
be an additional $\pm 1$--2\% systematic uncertainty.

\begin{table}[htb]
\caption{LO and NLO cross sections in (pb) for $p\bar p\to
W^{\prime}_{R,\,L}\to t\bar b + \bar t b$ at run I of the Tevatron,
$\sqrt{S}=1.8$ TeV, when the decay to leptons is not allowed.  Scale,
top-quark mass, and PDF uncertainties are also listed in (pb).  The
last column in each section lists the total theoretical uncertainty in
(\%), where all uncertainties in the text and in this Table are added
in quadrature.\label{tab:sigq}}
\begin{center}
\begin{tabular}{d|d@{\hspace*{2em}}l@{\hspace*{1em}}l@{\hspace*{1em}}l@{\hspace*{3em}}l@{\hspace*{1em}}|d@{\hspace*{2em}}l@{\hspace*{1em}}l@{\hspace*{1em}}l@{\hspace*{3em}}l}
Mass (GeV)&$\sigma_{\mathrm LO}$ (pb)&\multicolumn{3}{l}{$\delta\sigma_{\mathrm
LO}(\mu, \delta m_t$, PDF) (pb)}& $\delta\sigma_{\mathrm LO}^{\mathrm Tot}$ (\%)&
$\sigma_{\mathrm NLO}$ (pb)&\multicolumn{3}{l}{$\delta\sigma_{\mathrm NLO}(\mu, 
\delta m_t$, PDF) (pb)}& $\delta\sigma_{\mathrm NLO}^{\mathrm Tot}$ (\%) \\ \hline &&&&&&&&&&\\
200. & 32.24 & $^{+2.0}_{-1.8}$ & $^{+10.9}_{-9.8}$ & $^{+1.3}_{-1.2}$ & $^{+34.7}_{-31.2}$ & 50.90 & $^{+3.0}_{-2.4}$ & $^{+16}_{-15}$ & $^{+2.1}_{-1.9}$ & $^{+32.3}_{-30.2}$ \\ &&&&&&&&&&\\
225. & 51.45 & $^{+3.6}_{-3.2}$ & $^{+7.5}_{-7.4}$ & $^{+2.3}_{-2.0}$ & $^{+16.8}_{-16.2}$ & 77.24 & $^{+4.3}_{-3.6}$ & $^{+10}_{-10}$ & $^{+3.4}_{-3.0}$ & $^{+14.9}_{-14.4}$ \\ &&&&&&&&&&\\
250. & 52.29 & $^{+4.1}_{-3.6}$ & $^{+4.4}_{-4.5}$ & $^{+2.5}_{-2.1}$ & $^{+12.5}_{-11.8}$ & 76.10 & $^{+4.1}_{-3.6}$ & $^{+5.9}_{-6.1}$ & $^{+3.6}_{-3.0}$ & $^{+10.7}_{-10.2}$ \\ &&&&&&&&&&\\
275. & 45.58 & $^{+3.9}_{-3.4}$ & $^{+2.5}_{-2.6}$ & $^{+2.3}_{-1.9}$ & $^{+11.4}_{-10.3}$ & 64.92 & $^{+3.4}_{-3.1}$ & $^{+3.3}_{-3.4}$ & $^{+3.3}_{-2.7}$ & $^{+9.0}_{-8.4}$ \\ &&&&&&&&&&\\
300. & 37.18 & $^{+3.4}_{-2.9}$ & $^{+1.5}_{-1.6}$ & $^{+2.0}_{-1.6}$ & $^{+11.4}_{-9.9}$ & 52.11 & $^{+2.8}_{-2.6}$ & $^{+1.9}_{-2.0}$ & $^{+2.8}_{-2.2}$ & $^{+8.6}_{-7.7}$ \\ &&&&&&&&&&\\
350. & 22.82 & $^{+2.4}_{-2.0}$ & $^{+0.56}_{-0.58}$ & $^{+1.4}_{-1.1}$  & $^{+12.5}_{-10.4}$ & 31.20 & $^{+1.7}_{-1.6}$ & $^{+0.70}_{-0.73}$ & $^{+1.9}_{-1.4}$ & $^{+8.6}_{-7.4}$ \\ &&&&&&&&&&\\
400. & 13.46 & $^{+1.5}_{-1.3}$ & $^{+0.23}_{-0.23}$ & $^{+0.91}_{-0.68}$ & $^{+13.2}_{-11.1}$ & 18.04 & $^{+1.03}_{-0.98}$ & $^{+0.28}_{-0.29}$ & $^{+1.22}_{-0.91}$ & $^{+9.1}_{-7.7}$ \\ &&&&&&&&&&\\
450. & 7.836 & $^{+0.98}_{-0.82}$ & $^{+0.10}_{-0.10}$ & $^{+0.59}_{-0.44}$ & $^{+14.7}_{-12.0}$ & 10.31 & $^{+0.62}_{-0.59}$ & $^{+0.12}_{-0.12}$ & $^{+0.78}_{-0.58}$ & $^{+9.8}_{-8.2}$ \\ &&&&&&&&&&\\
500. & 4.537 & $^{+0.61}_{-0.51}$ & $^{+0.045}_{-0.046}$ & $^{+0.39}_{-0.28}$ & $^{+16.0}_{-12.9}$ & 5.873 & $^{+0.38}_{-0.36}$ & $^{+0.054}_{-0.056}$ & $^{+0.50}_{-0.37}$ & $^{+10.8}_{-9.0}$ \\ &&&&&&&&&&\\
550. & 2.617 & $^{+0.38}_{-0.31}$ & $^{+0.022}_{-0.022}$ & $^{+0.26}_{-0.18}$ & $^{+17.6}_{-13.8}$ & 3.335 & $^{+0.23}_{-0.21}$ & $^{+0.026}_{-0.027}$ & $^{+0.33}_{-0.23}$ & $^{+12.2}_{-9.5}$ \\ &&&&&&&&&&\\
600. & 1.503 & $^{+0.23}_{-0.19}$ & $^{+0.011}_{-0.011}$ & $^{+0.17}_{-0.12}$ & $^{+19.1}_{-15.0}$ & 1.889 & $^{+0.14}_{-0.13}$ & $^{+0.013}_{-0.013}$ & $^{+0.21}_{-0.15}$ & $^{+13.4}_{-10.6}$ \\ &&&&&&&&&&\\
650. & 0.859 & $^{+0.14}_{-0.11}$ & $^{+0.006}_{-0.006}$ & $^{+0.11}_{-0.07}$ & $^{+20.8}_{-15.2}$ & 1.066 & $^{+0.08}_{-0.08}$ & $^{+0.007}_{-0.007}$ & $^{+0.13}_{-0.09}$ & $^{+14.4}_{-11.4}$
\end{tabular}
\end{center}
\end{table}

\begin{table}[htb]
\caption{LO and NLO cross sections in (pb) for $p\bar p\to
W^{\prime}_{R,\,L} \to t\bar b + \bar t b$ at run I of the Tevatron,
$\sqrt{S}=1.8$ TeV, when decays to quarks or leptons are both
included.  Scale, top-quark mass, and PDF uncertainties are also
listed in (pb).  The last column in each section lists the total
theoretical uncertainty in (\%), where all uncertainties in the text
and in this Table are added in quadrature.\label{tab:sigql}}
\begin{center}
\begin{tabular}{d|d@{\hspace*{2em}}l@{\hspace*{1em}}l@{\hspace*{1em}}l@{\hspace*{3em}}l@{\hspace*{1em}}|d@{\hspace*{2em}}l@{\hspace*{1em}}l@{\hspace*{1em}}l@{\hspace*{3em}}l}
Mass (GeV)&$\sigma_{\mathrm LO}$ (pb)&\multicolumn{3}{l}{$\delta\sigma_{\mathrm
LO}(\mu, \delta m_t$, PDF) (pb)}& $\delta\sigma_{\mathrm LO}^{\mathrm Tot}$ (\%)&
$\sigma_{\mathrm NLO}$ (pb)&\multicolumn{3}{l}{$\delta\sigma_{\mathrm NLO}(\mu, 
\delta m_t$, PDF) (pb)}& $\delta\sigma_{\mathrm NLO}^{\mathrm Tot}$ (\%) \\ \hline &&&&&&&&&&\\
200. & 21.91 & $^{+1.4}_{-1.3}$ & $^{+7.3}_{-6.5}$ & $^{+0.91}_{-0.84}$ & $^{+34.2}_{-30.5}$ & 34.94 & $^{+2.0}_{-1.7}$ & $^{+11.1}_{-10.1}$ & $^{+1.5}_{-1.3}$ & $^{+32.7}_{-29.6}$ \\ &&&&&&&&&&\\
225. & 35.05 & $^{+2.5}_{-2.2}$ & $^{+5.2}_{-5.1}$ & $^{+1.5}_{-1.4}$ & $^{+17.0}_{-16.4}$ & 53.36 & $^{+2.9}_{-2.5}$ & $^{+7.5}_{-7.4}$ & $^{+2.4}_{-2.1}$ & $^{+15.8}_{-15.3}$ \\ &&&&&&&&&&\\
250. & 36.15 & $^{+2.8}_{-2.5}$ & $^{+3.2}_{-3.2}$ & $^{+1.7}_{-1.4}$ & $^{+12.7}_{-11.9}$ & 53.40 & $^{+2.9}_{-2.5}$ & $^{+4.4}_{-4.5}$ & $^{+2.5}_{-2.1}$ & $^{+11.1}_{-10.6}$ \\ &&&&&&&&&&\\
275. & 31.92 & $^{+2.7}_{-2.4}$ & $^{+1.9}_{-1.9}$ & $^{+1.6}_{-1.3}$ & $^{+11.5}_{-10.5}$ & 46.14 & $^{+2.4}_{-2.2}$ & $^{+2.5}_{-2.6}$ & $^{+2.3}_{-1.9}$ & $^{+9.2}_{-8.6}$ \\ &&&&&&&&&&\\
300. & 26.31 & $^{+2.4}_{-2.1}$ & $^{+1.2}_{-1.2}$ & $^{+1.4}_{-1.1}$ & $^{+11.5}_{-10.1}$ & 37.41 & $^{+2.0}_{-1.8}$ & $^{+1.5}_{-1.6}$ & $^{+2.0}_{-1.6}$ & $^{+8.7}_{-7.9}$ \\ &&&&&&&&&&\\
350. & 16.40 & $^{+1.7}_{-1.4}$ & $^{+0.44}_{-0.46}$ & $^{+0.98}_{-0.75}$ & $^{+12.3}_{-10.1}$ & 22.72 & $^{+1.2}_{-1.2}$ & $^{+0.56}_{-0.58}$ & $^{+1.4}_{-1.0}$ & $^{+8.6}_{-7.5}$ \\ &&&&&&&&&&\\
400. & 9.773 & $^{+1.12}_{-0.95}$ & $^{+0.18}_{-0.19}$ & $^{+0.66}_{-0.49}$ & $^{+13.5}_{-11.2}$ & 13.26 & $^{+0.75}_{-0.72}$ & $^{+0.23}_{-0.24}$ & $^{+0.89}_{-0.67}$ & $^{+9.1}_{-7.8}$ \\ &&&&&&&&&&\\
450. & 5.735 & $^{+0.72}_{-0.60}$ & $^{+0.081}_{-0.083}$ & $^{+0.44}_{-0.32}$ & $^{+14.8}_{-12.0}$ & 7.641 & $^{+0.46}_{-0.44}$ & $^{+0.10}_{-0.10}$ & $^{+0.58}_{-0.43}$ & $^{+9.9}_{-8.3}$ \\ &&&&&&&&&&\\
500. & 3.343 & $^{+0.45}_{-0.37}$ & $^{+0.038}_{-0.039}$ & $^{+0.29}_{-0.21}$ & $^{+16.1}_{-12.8}$ & 4.380 & $^{+0.28}_{-0.27}$ & $^{+0.046}_{-0.047}$ & $^{+0.38}_{-0.27}$ & $^{+10.9}_{-8.9}$ \\ &&&&&&&&&&\\
550. & 1.941 & $^{+0.28}_{-0.23}$ & $^{+0.018}_{-0.019}$ & $^{+0.19}_{-0.14}$ & $^{+17.5}_{-13.9}$ & 2.503 & $^{+0.17}_{-0.16}$ & $^{+0.022}_{-0.023}$ & $^{+0.24}_{-0.17}$ & $^{+11.9}_{-9.5}$ \\ &&&&&&&&&&\\
600. & 1.122 & $^{+0.17}_{-0.14}$ & $^{+0.009}_{-0.009}$ & $^{+0.12}_{-0.09}$ & $^{+18.6}_{-14.9}$ & 1.427 & $^{+0.10}_{-0.10}$ & $^{+0.011}_{-0.011}$ & $^{+0.16}_{-0.11}$ & $^{+13.3}_{-10.5}$ \\ &&&&&&&&&&\\
650. & 0.646 & $^{+0.11}_{-0.08}$ & $^{+0.005}_{-0.005}$ & $^{+0.08}_{-0.06}$ & $^{+21.1}_{-15.5}$ & 0.812 & $^{+0.06}_{-0.06}$ & $^{+0.006}_{-0.006}$ & $^{+0.10}_{-0.07}$ & $^{+14.4}_{-11.5}$
\end{tabular}
\end{center}
\end{table}

\begin{table}[htb]
\caption{LO and NLO cross sections in (pb) for $p\bar p\to
W^{\prime}_{R,\,L} \to t\bar b + \bar t b$ at run II of the Tevatron,
$\sqrt{S}=1.96$ TeV, when the decay to leptons is not allowed.  Scale,
top-quark mass, and PDF uncertainties are also listed in (pb).  The
last column in each section lists the total theoretical uncertainty in
(\%), where all uncertainties in the text and in this Table are added
in quadrature.\label{tab:sigqt}}
\begin{center}
\begin{tabular}{d|d@{\hspace*{2em}}l@{\hspace*{1em}}l@{\hspace*{1em}}l@{\hspace*{3em}}l@{\hspace*{1em}}|d@{\hspace*{2em}}l@{\hspace*{1em}}l@{\hspace*{1em}}l@{\hspace*{3em}}l}
Mass (GeV)&$\sigma_{\mathrm LO}$ (pb)&\multicolumn{3}{l}{$\delta\sigma_{\mathrm
LO}(\mu, \delta m_t$, PDF) (pb)}& $\delta\sigma_{\mathrm LO}^{\mathrm Tot}$ (\%)&
$\sigma_{\mathrm NLO}$ (pb)&\multicolumn{3}{l}{$\delta\sigma_{\mathrm NLO}(\mu, 
\delta m_t$, PDF) (pb)}& $\delta\sigma_{\mathrm NLO}^{\mathrm Tot}$ (\%) \\ \hline &&&&&&&&&&\\
200. & 36.39 & $^{+2.1}_{-1.9}$ & $^{+12.3}_{-11.0}$ & $^{+1.5}_{-1.4}$ & $^{+34.6}_{-31.0}$ & 57.39 & $^{+3.2}_{-2.6}$ & $^{+18.4}_{-16.8}$ & $^{+2.3}_{-2.2}$ & $^{+33.1}_{-30.2}$ \\ &&&&&&&&&& \\
225. & 58.56 & $^{+3.8}_{-3.4}$ & $^{+8.5}_{-8.4}$ & $^{+2.5}_{-2.2}$ & $^{+16.5}_{-15.9}$ & 87.86 & $^{+4.6}_{-3.9}$ & $^{+11.8}_{-11.9}$ & $^{+3.7}_{-3.4}$ & $^{+15.6}_{-15.3}$ \\ &&&&&&&&&& \\
250. & 60.13 & $^{+4.3}_{-3.8}$ & $^{+5.0}_{-5.1}$ & $^{+2.7}_{-2.3}$ & $^{+11.9}_{-11.2}$ & 87.52 & $^{+4.5}_{-3.9}$ & $^{+6.8}_{-6.9}$ & $^{+3.9}_{-3.4}$ & $^{+11.1}_{-10.7}$ \\ &&&&&&&&&& \\
275. & 53.00 & $^{+4.1}_{-3.6}$ & $^{+3.0}_{-3.0}$ & $^{+2.5}_{-2.1}$ & $^{+10.7}_{-9.7}$ & 75.57 & $^{+3.8}_{-3.4}$ & $^{+3.9}_{-4.0}$ & $^{+3.6}_{-3.0}$ & $^{+9.6}_{-9.0}$ \\ &&&&&&&&&& \\
300. & 43.78 & $^{+3.7}_{-3.2}$ & $^{+1.8}_{-1.8}$ & $^{+2.2}_{-1.8}$ & $^{+10.7}_{-9.3}$ & 61.46 & $^{+3.1}_{-2.8}$ & $^{+2.3}_{-2.4}$ & $^{+3.1}_{-2.5}$ & $^{+9.0}_{-8.3}$ \\ &&&&&&&&&& \\
350. & 27.63 & $^{+2.6}_{-2.3}$ & $^{+0.68}_{-0.70}$ & $^{+1.6}_{-1.2}$ & $^{+11.3}_{-9.7}$ & 37.90 & $^{+1.9}_{-1.8}$ & $^{+0.85}_{-0.88}$ & $^{+2.1}_{-1.7}$ & $^{+8.8}_{-8.1}$ \\ &&&&&&&&&& \\
400. & 16.81 & $^{+1.8}_{-1.5}$ & $^{+0.28}_{-0.29}$ & $^{+1.05}_{-0.80}$ & $^{+12.5}_{-10.3}$ & 22.65 & $^{+1.2}_{-1.1}$ & $^{+0.35}_{-0.36}$ & $^{+1.4}_{-1.1}$ & $^{+9.3}_{-8.2}$ \\ &&&&&&&&&& \\
450. & 10.13 & $^{+1.2}_{-1.0}$ & $^{+0.13}_{-0.13}$ & $^{+0.70}_{-0.52}$ & $^{+13.8}_{-11.2}$ & 13.43 & $^{+0.76}_{-0.72}$ & $^{+0.15}_{-0.16}$ & $^{+0.93}_{-0.70}$ & $^{+9.9}_{-8.6}$ \\ &&&&&&&&&& \\
500. & 6.092 & $^{+0.76}_{-0.64}$ & $^{+0.060}_{-0.061}$ & $^{+0.47}_{-0.35}$ & $^{+14.7}_{-12.0}$ & 7.953 & $^{+0.47}_{-0.45}$ & $^{+0.072}_{-0.074}$ & $^{+0.62}_{-0.45}$ & $^{+10.7}_{-9.0}$ \\ &&&&&&&&&& \\
550. & 3.661 & $^{+0.49}_{-0.41}$ & $^{+0.030}_{-0.030}$ & $^{+0.32}_{-0.23}$ & $^{+16.0}_{-12.9}$ & 4.710 & $^{+0.30}_{-0.28}$ & $^{+0.035}_{-0.036}$ & $^{+0.41}_{-0.30}$ & $^{+11.6}_{-9.7}$ \\ &&&&&&&&&& \\
600. & 2.197 & $^{+0.32}_{-0.26}$ & $^{+0.015}_{-0.016}$ & $^{+0.21}_{-0.15}$ & $^{+17.4}_{-13.7}$ & 2.790 & $^{+0.18}_{-0.18}$ & $^{+0.018}_{-0.019}$ & $^{+0.27}_{-0.19}$ & $^{+12.4}_{-10.3}$ \\ &&&&&&&&&& \\
650. & 1.316 & $^{+0.20}_{-0.16}$ & $^{+0.0081}_{-0.0082}$ & $^{+0.14}_{-0.10}$ & $^{+18.6}_{-14.4}$ & 1.650 & $^{+0.11}_{-0.11}$ & $^{+0.010}_{-0.010}$ & $^{+0.18}_{-0.13}$ & $^{+13.4}_{-11.1}$ \\ &&&&&&&&&& \\
700. & 0.7855 & $^{+0.13}_{-0.10}$ & $^{+0.0045}_{-0.0045}$ & $^{+0.096}_{-0.067}$ & $^{+20.6}_{-15.3}$ & 0.974 & $^{+0.071}_{-0.069}$ & $^{+0.0054}_{-0.0055}$ & $^{+0.12}_{-0.08}$ & $^{+14.9}_{-11.6}$ \\ &&&&&&&&&& \\
750. & 0.4671 & $^{+0.080}_{-0.063}$ & $^{+0.0026}_{-0.0026}$ & $^{+0.063}_{-0.044}$ & $^{+21.8}_{-16.5}$ & 0.573 & $^{+0.043}_{-0.043}$ & $^{+0.0032}_{-0.0031}$ & $^{+0.077}_{-0.054}$ & $^{+15.9}_{-12.7}$ \\ &&&&&&&&&& \\
800. & 0.2767 & $^{+0.049}_{-0.039}$ & $^{+0.0015}_{-0.0015}$ & $^{+0.041}_{-0.028}$ & $^{+23.1}_{-17.4}$ & 0.337 & $^{+0.027}_{-0.027}$ & $^{+0.0019}_{-0.0019}$ & $^{+0.050}_{-0.034}$ & $^{+17.4}_{-13.5}$ \\ &&&&&&&&&& \\
850. & 0.1634 & $^{+0.030}_{-0.024}$ & $^{+0.0010}_{-0.0010}$ & $^{+0.026}_{-0.018}$ & $^{+24.3}_{-18.4}$ & 0.198 & $^{+0.016}_{-0.016}$ & $^{+0.0012}_{-0.0012}$ & $^{+0.032}_{-0.022}$ & $^{+18.5}_{-14.4}$ \\ &&&&&&&&&& \\
900. & 0.0963 & $^{+0.018}_{-0.014}$ & $^{+0.0006}_{-0.0006}$ & $^{+0.017}_{-0.011}$ & $^{+25.7}_{-18.5}$ & 0.116 & $^{+0.010}_{-0.010}$ & $^{+0.0008}_{-0.0008}$ & $^{+0.020}_{-0.014}$ & $^{+19.7}_{-15.4}$ \\ &&&&&&&&&& \\
950. & 0.0569 & $^{+0.011}_{-0.009}$ & $^{+0.0004}_{-0.0004}$ & $^{+0.010}_{-0.007}$ & $^{+26.1}_{-20.1}$ & 0.0688 & $^{+0.0059}_{-0.0060}$ & $^{+0.0006}_{-0.0006}$ & $^{+0.0127}_{-0.0084}$ & $^{+20.8}_{-15.6}$ \\ &&&&&&&&&& \\
1000.& 0.0339 & $^{+0.006}_{-0.005}$ & $^{+0.0003}_{-0.0003}$ & $^{+0.0064}_{-0.0042}$ & $^{+25.9}_{-19.3}$ & 0.0413 & $^{+0.0036}_{-0.0036}$ & $^{+0.0004}_{-0.0004}$ & $^{+0.0079}_{-0.0051}$ & $^{+21.4}_{-15.7}$
\end{tabular}
\end{center}
\end{table}

\begin{table}[htb]
\caption{LO and NLO cross sections in (pb) for $p\bar p\to
W^{\prime}_{R,\,L} \to t\bar b + \bar t b$ at run II of the Tevatron,
$\sqrt{S}=1.96$ TeV, when decays to quarks or leptons are both
included.  Scale, top-quark mass, and PDF uncertainties are also
listed in (pb).  The last column in each section lists the total
theoretical uncertainty in (\%), where all uncertainties in the text
and in this Table are added in quadrature.\label{tab:sigqlt}}
\begin{center}
\begin{tabular}{d|d@{\hspace*{2em}}l@{\hspace*{1em}}l@{\hspace*{1em}}l@{\hspace*{3em}}l@{\hspace*{1em}}|d@{\hspace*{2em}}l@{\hspace*{1em}}l@{\hspace*{1em}}l@{\hspace*{3em}}l}
Mass (GeV)&$\sigma_{\mathrm LO}$ (pb)&\multicolumn{3}{l}{$\delta\sigma_{\mathrm
LO}(\mu, \delta m_t$, PDF) (pb)}& $\delta\sigma_{\mathrm LO}^{\mathrm Tot}$ (\%)&
$\sigma_{\mathrm NLO}$ (pb)&\multicolumn{3}{l}{$\delta\sigma_{\mathrm NLO}(\mu, 
\delta m_t$, PDF) (pb)}& $\delta\sigma_{\mathrm NLO}^{\mathrm Tot}$ (\%) \\ \hline &&&&&&&&&&\\
200. & 24.75 & $^{+1.4}_{-1.3}$ & $^{+8.3}_{-7.3}$ & $^{+0.99}_{-0.94}$ & $^{+34.6}_{-31.0}$ & 39.43 & $^{+2.2}_{-1.8}$ & $^{+12.5}_{-11.3}$ & $^{+1.6}_{-1.5}$ & $^{+32.7}_{-29.6}$ \\ &&&&&&&&&& \\
225. & 39.92 & $^{+2.6}_{-2.3}$ & $^{+6.0}_{-5.8}$ & $^{+1.7}_{-1.5}$ & $^{+16.5}_{-15.9}$ & 60.74 & $^{+3.2}_{-2.7}$ & $^{+8.5}_{-8.4}$ & $^{+2.6}_{-2.3}$ & $^{+16.1}_{-15.6}$ \\ &&&&&&&&&& \\
250. & 41.58 & $^{+3.0}_{-2.6}$ & $^{+3.7}_{-3.7}$ & $^{+1.9}_{-1.6}$ & $^{+11.9}_{-11.2}$ & 61.43 & $^{+3.1}_{-2.7}$ & $^{+5.0}_{-5.1}$ & $^{+2.7}_{-2.4}$ & $^{+11.3}_{-11.0}$ \\ &&&&&&&&&& \\
275. & 37.13 & $^{+2.9}_{-2.5}$ & $^{+2.2}_{-2.3}$ & $^{+1.8}_{-1.5}$ & $^{+10.7}_{-9.7}$ & 53.73 & $^{+2.7}_{-2.4}$ & $^{+3.0}_{-3.0}$ & $^{+2.5}_{-2.1}$ & $^{+9.8}_{-9.1}$ \\ &&&&&&&&&& \\
300. & 30.99 & $^{+2.6}_{-2.3}$ & $^{+1.4}_{-1.4}$ & $^{+1.6}_{-1.3}$ & $^{+10.7}_{-9.3}$ & 44.14 & $^{+2.2}_{-2.0}$ & $^{+1.8}_{-1.8}$ & $^{+2.2}_{-1.8}$ & $^{+9.1}_{-8.4}$ \\ &&&&&&&&&& \\
350. & 19.85 & $^{+1.9}_{-1.6}$ & $^{+0.54}_{-0.55}$ & $^{+1.11}_{-0.87}$ & $^{+11.3}_{-9.7}$ & 27.61 & $^{+1.4}_{-1.3}$ & $^{+0.68}_{-0.70}$ & $^{+1.5}_{-1.2}$ & $^{+8.8}_{-8.0}$ \\ &&&&&&&&&& \\
400. & 12.20 & $^{+1.3}_{-1.1}$ & $^{+0.23}_{-0.23}$ & $^{+0.76}_{-0.58}$ & $^{+12.5}_{-10.3}$ & 16.65 & $^{+0.88}_{-0.84}$ & $^{+0.28}_{-0.29}$ & $^{+1.04}_{-0.79}$ & $^{+9.3}_{-8.2}$ \\ &&&&&&&&&& \\
450. & 7.411 & $^{+0.86}_{-0.73}$ & $^{+0.10}_{-0.11}$ & $^{+0.52}_{-0.38}$ & $^{+13.8}_{-11.2}$ & 9.941 & $^{+0.56}_{-0.53}$ & $^{+0.13}_{-0.13}$ & $^{+0.69}_{-0.51}$ & $^{+9.9}_{-8.6}$ \\ &&&&&&&&&& \\
500. & 4.484 & $^{+0.56}_{-0.47}$ & $^{+0.050}_{-0.051}$ & $^{+0.35}_{-0.26}$ & $^{+14.7}_{-12.0}$ & 5.922 & $^{+0.35}_{-0.33}$ & $^{+0.061}_{-0.062}$ & $^{+0.46}_{-0.34}$ & $^{+10.6}_{-9.1}$ \\ &&&&&&&&&& \\
550. & 2.709 & $^{+0.36}_{-0.30}$ & $^{+0.025}_{-0.026}$ & $^{+0.24}_{-0.17}$ & $^{+16.0}_{-12.9}$ & 3.526 & $^{+0.22}_{-0.21}$ & $^{+0.030}_{-0.031}$ & $^{+0.31}_{-0.22}$ & $^{+11.6}_{-9.6}$ \\ &&&&&&&&&& \\
600. & 1.635 & $^{+0.23}_{-0.19}$ & $^{+0.013}_{-0.013}$ & $^{+0.16}_{-0.11}$ & $^{+17.4}_{-13.7}$ & 2.100 & $^{+0.14}_{-0.13}$ & $^{+0.016}_{-0.016}$ & $^{+0.20}_{-0.15}$ & $^{+12.4}_{-10.3}$ \\ &&&&&&&&&& \\
650. & 0.9849 & $^{+0.15}_{-0.12}$ & $^{+0.0071}_{-0.0072}$ & $^{+0.108}_{-0.076}$ & $^{+18.6}_{-14.4}$ & 1.250 & $^{+0.086}_{-0.084}$ & $^{+0.0085}_{-0.0087}$ & $^{+0.14}_{-0.10}$ & $^{+13.8}_{-11.3}$ \\ &&&&&&&&&& \\
700. & 0.5919 & $^{+0.095}_{-0.076}$ & $^{+0.0040}_{-0.0040}$ & $^{+0.072}_{-0.050}$ & $^{+20.6}_{-15.3}$ & 0.7428 & $^{+0.054}_{-0.053}$ & $^{+0.0048}_{-0.0049}$ & $^{+0.090}_{-0.063}$ & $^{+14.7}_{-11.8}$ \\ &&&&&&&&&& \\
750. & 0.3547 & $^{+0.060}_{-0.047}$ & $^{+0.0023}_{-0.0023}$ & $^{+0.048}_{-0.033}$ & $^{+21.8}_{-16.5}$ & 0.4410 & $^{+0.033}_{-0.033}$ & $^{+0.0029}_{-0.0029}$ & $^{+0.059}_{-0.041}$ & $^{+15.9}_{-12.6}$ \\ &&&&&&&&&& \\
800. & 0.2122 & $^{+0.037}_{-0.029}$ & $^{+0.0014}_{-0.0014}$ & $^{+0.031}_{-0.022}$ & $^{+23.1}_{-17.4}$ & 0.2619 & $^{+0.020}_{-0.020}$ & $^{+0.0018}_{-0.0018}$ & $^{+0.039}_{-0.027}$ & $^{+17.2}_{-13.5}$ \\ &&&&&&&&&& \\
850. & 0.1268 & $^{+0.023}_{-0.018}$ & $^{+0.0009}_{-0.0009}$ & $^{+0.020}_{-0.014}$ & $^{+24.3}_{-18.4}$ & 0.1558 & $^{+0.013}_{-0.013}$ & $^{+0.0012}_{-0.0012}$ & $^{+0.025}_{-0.017}$ & $^{+18.6}_{-14.4}$ \\ &&&&&&&&&& \\
900. & 0.0759 & $^{+0.014}_{-0.011}$ & $^{+0.0006}_{-0.0006}$ & $^{+0.0132}_{-0.0089}$ & $^{+25.7}_{-18.5}$ & 0.0932 & $^{+0.0077}_{-0.0077}$ & $^{+0.0008}_{-0.0008}$ & $^{+0.016}_{-0.011}$ & $^{+19.5}_{-15.0}$ \\ &&&&&&&&&& \\
950. & 0.0457 & $^{+0.009}_{-0.007}$ & $^{+0.0004}_{-0.0004}$ & $^{+0.0084}_{-0.0056}$ & $^{+26.1}_{-20.1}$ & 0.0564 & $^{+0.0047}_{-0.0047}$ & $^{+0.0006}_{-0.0006}$ & $^{+0.0104}_{-0.0069}$ & $^{+20.7}_{-15.4}$ \\ &&&&&&&&&& \\
1000.& 0.0279 & $^{+0.005}_{-0.004}$ & $^{+0.0003}_{-0.0003}$ & $^{+0.0053}_{-0.0035}$ & $^{+25.9}_{-19.3}$ & 0.0348 & $^{+0.0029}_{-0.0029}$ & $^{+0.0004}_{-0.0004}$ & $^{+0.0066}_{-0.0043}$ & $^{+21.2}_{-15.5}$
\end{tabular}
\end{center}
\end{table}

\begin{figure}[tbh]
\epsfxsize=232pt
\centerline{\epsfbox{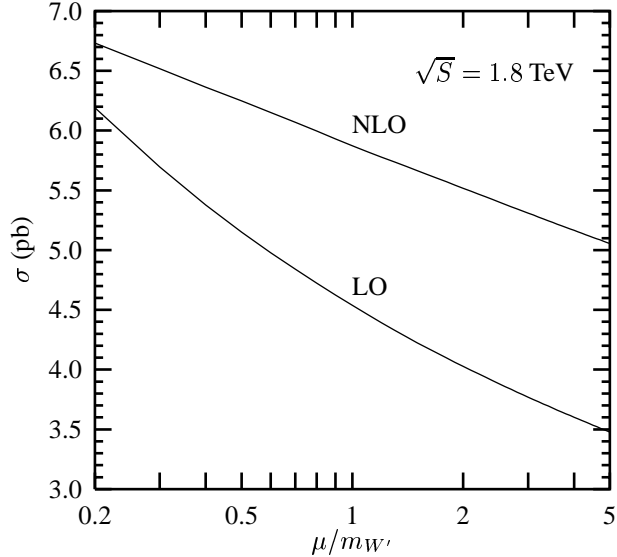}} % 232pt actual size
\caption{Scale dependence of the leading-order (LO) and
next-to-leading order (NLO) cross section $p\bar p\to \wprm \to t\bar
b,\bar t b$ for $m_{W^{\prime}}=500$ GeV at run I of the Tevatron
($\sqrt{S}=1.8$~TeV).\label{fig:muerr}}
\end{figure}

In Tables~\ref{tab:sigq}--\ref{tab:sigqlt} we use $m_{\wprm}/2$ and
$2m_{\wprm}$ to estimate the uncertainty due to the choice of scale
$\mu$.  While this is a reasonable estimate, we show in
Fig.~\ref{fig:muerr} the scale dependence of the LO and NLO cross
section for a typical mass $m_{W^{\prime}}=500$ GeV at run I of the
Tevatron over the range $\mu = m_{\wprm}/5$--$5m_{\wprm}$.  Even
though the scale only enters the LO cross section though the PDFs, the
NLO cross section is much less scale dependent than the LO cross
section.  It is apparent from this figure that only at unnaturally
small scales do the LO and NLO cross sections agree.  This is typical
of single-top-quark production \cite{Harris:2002md,Smith:1996ij}, and
Drell-Yan-like processes in general, where initial-state corrections
act like new production modes.

\begin{figure}[tbh]
\epsfxsize=240pt
\centerline{\epsfbox{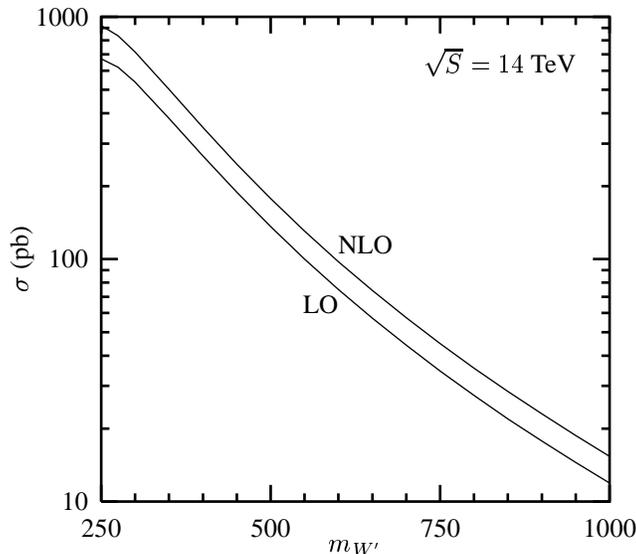}} % 240pt actual size
\caption{Leading-order (LO) and next-to-leading order (NLO) cross
section $p p\to W^{\prime}_{R,\,L} \to t\bar b,\bar t b$ as a function
of $m_{W^{\prime}}$ at the LHC ($\sqrt{S}=14$~TeV), when the decay to
leptons is not allowed.\label{fig:siglhc}}
\end{figure}

\begin{figure}[tbh]
\epsfxsize=240pt
\centerline{\epsfbox{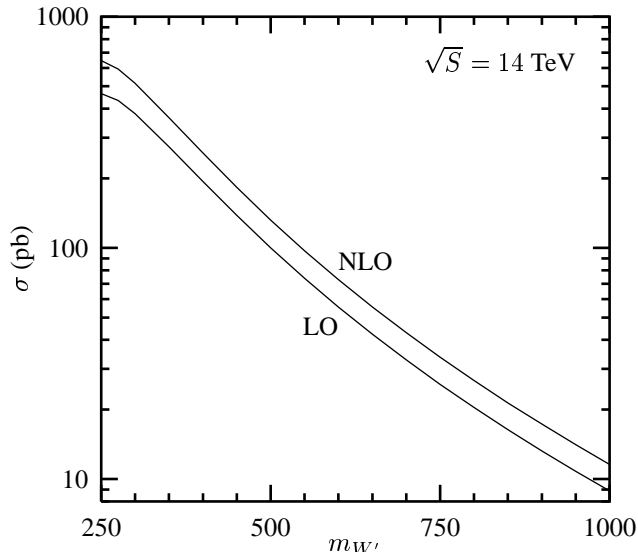}} % 240pt actual size
\caption{Leading-order (LO) and next-to-leading order (NLO) cross
section $p p\to W^{\prime}_{R,\,L} \to t\bar b,\bar t b$ as a function
of $m_{W^{\prime}}$ at the LHC ($\sqrt{S}=14$~TeV), when decays to
quarks and leptons are both included.\label{fig:siglhcl}}
\end{figure}

In Figs.~\ref{fig:siglhc} and \ref{fig:siglhcl}, we show the LO and
NLO single-top-quark cross sections, with $W^\prime_{R,\,L}$ bosons,
at the LHC as a function of $m_{\wprm}$.  The cross sections are
roughly a factor of 30 larger at the LHC than at the Tevatron due to
the larger quark luminosity.  For \wpr masses above 1~TeV, the cross
section as a function of mass decreases slowly, leading to a much
larger mass reach at the LHC.  For example, the cross sections where
quark and lepton decays are allowed are 580, 62, 9, and 2~fb, for 2,
3, 4, and 5~TeV \wpr bosons, respectively.  Depending on detector
performance, limits of 3--4~TeV on the \wpr mass should appear within
a year or two of running at the initial luminosity of 10~fb$^{-1}$ per
year.

\subsection{Parton distribution function uncertainties}
\label{sec:pdfu}

Because we are interested in heavy particles, the parton distribution
functions (PDFs) will be probed at large values of the proton's
momentum fraction.  Hence, as opposed to Drell-Yan or $s$-channel
single-top-quark production where the PDFs are well understood, we
expect the PDF uncertainty to play a significant role in the total
uncertainty of the cross section.  In order to estimate the effects of
this uncertainty we use a modification
\cite{Sullivan:2001ry,Nadolsky:2001yg} of the ``tolerance method''
implemented in the CTEQ6 PDFs \cite{Pumplin:2002vw}.

The tolerance method is based on diagonalization of the matrix of
second derivatives for $\chi^{2}$ (a Hessian matrix) near the minimum
of $\chi^{2}$ for the PDF fits \cite{Pumplin:2001ct}.  Since
$\chi^{2}$ is approximately parabolic near its minimum $\chi^{2}_{0}$,
hypersurfaces of constant $\chi^{2}$ are hyperellipses in the space of
the original 20 PDF parameters $\{a_{i}\}$.  These hyperellipses are
transformed into hyperspheres by a change of coordinates
$\{a_{i}\}\rightarrow \{z_{i}\},\, i=1,\dots ,20$.  The tolerance
method assumes that all acceptable PDF sets correspond to a $\chi^{2}$
that does not exceed their minimal value $\chi _{0}^{2}$ more than by
$T^{2}$.  As a result, the acceptable PDF sets have $\{z_{i}\}$ within
a sphere of the radius $T^{2}$ around $\{z_{i}(\chi _{0}^{2})\}\equiv
\{z_{i}^{0}\}$.  In principle, $T$ should be chosen to correspond to a
$1\sigma$ deviation of the fit.  However, for simplicity we present
results for $T=10$, as used by the CTEQ6M101 -- CTEQ6M140 PDF tables
given in Ref. \cite{Pumplin:2002vw}.

The PDF uncertainty for an observable $O$ is the maximal change in $O$
as a function of variables $\{z_{i}\}$ varying within the tolerance
hypersphere.  The CTEQ6 paper estimates the variation of $O$ by using
a master formula
\begin{equation}
\label{eq:tolmast}
\delta O=\frac{1}{2}\sqrt{\sum^{20}_{i=1}\delta O_{i}^{2}},\mbox {\,
where\, }\delta O_{i}\equiv T\frac{\partial O}{\partial z_{i}}\approx
T\frac{O(z^{0}_{i}+t)-O(z^{0}_{i}-t)}{t} \;,
\end{equation}
and $t=10$ is a step in the space of $z_{i}$.  Here, $O(z_{1}^{0},\dots
,z_{i}^{0}\pm t,\dots ,z_{20}^{0})$ is denoted as $O(z_{i}^{0}\pm t)$.
Eq.~(\ref{eq:tolmast}) is a good approximation for the PDF fits, but is
less useful for observables, e.g.\ cross sections.

\begin{figure}[tbh]
\epsfxsize=3.4in
\centerline{\epsfbox{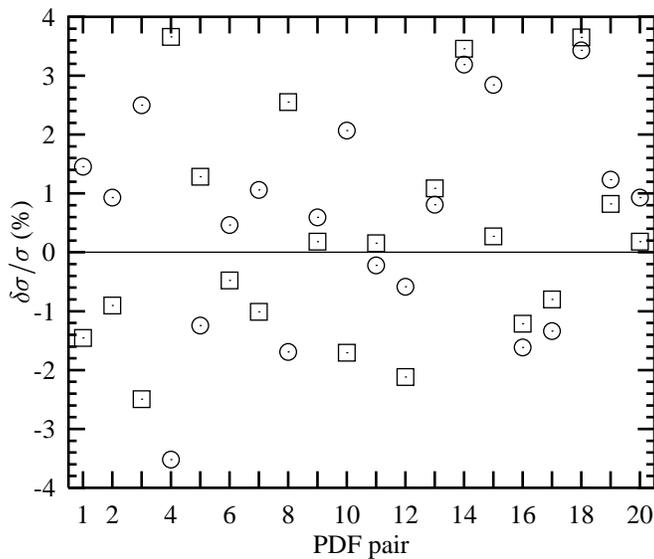}} % 246pt is natural size
\caption{Uncertainty in the cross section for $m_{\wprm} = 500$ GeV
for each pair of PDFs using CTEQ6M1xx.  Circles are odd PDF sets.
Squares are even PDF sets.\label{fig:errasm}}
\end{figure}

The difficulty with Eq.~(\ref{eq:tolmast}) is well exemplified for the
problem at hand by Fig.~\ref{fig:errasm}.  Here we see the uncertainty
for each of the 20 pairs of PDFs as a function of the parameters $z_i$
for $m_{\wprm} = 500$ GeV.  If we apply Eq.~(\ref{eq:tolmast}) we would
predict an uncertainty of $\pm 5.9\%$.  However,
$O(z^{0}_{i}+t)-O(z^{0}_{i}-t) \approx 0$ for nearly half of the
parameters, even though Fig.~\ref{fig:errasm} shows that these provide
a large deviation.  The uncertainty in Eq.~(\ref{eq:tolmast}) appears to
be more sensitive to $z_i$ which have a small absolute effect, but
happen to cause a shift in the prediction which changes sign, than to
large fluctuations.  This defect of the master formula is a result of
the simple observation that the PDF set that minimizes the uncertainty
in a given observable $O$ is not necessarily the same as the one that
minimizes the fit to the global data set.

In order to obtain a better estimator of the uncertainty of a generic
observable $O$, we introduce \cite{Sullivan:2001ry,Nadolsky:2001yg} the
``modified tolerance method'' (MTM) master formula.  We define the maximum
positive and negative errors on an observable $O$ by
\begin{eqnarray}
\delta O_{+} & = & \frac{T}{t}\sqrt{\sum_{i=1}^{20}\Bigl( \max 
[\,O(z^{0}_{i}+t)-O(z_{i}^{0}),O(z^{0}_{i}-t)-O(z_{i}^{0}),0]\Bigr)^2}
\;,\label{eq:pdferra} \\
\delta O_{-} & = & \frac{T}{t}\sqrt{\sum_{i=1}^{20}\Bigl( \max 
[\,O(z_{i}^{0})-O(z^{0}_{i}+t),O(z_{i}^{0})-O(z^{0}_{i}-t),0]\Bigr)^2}
\; ,\label{eq:pdferr}
\end{eqnarray}
where the ``tolerance'' $T$ is the same scaling parameter that
determines the overall range of allowed variation of $\chi^2$, and we
use $T=t=10$ as in the CTEQ6 PDF sets.  In
Eqs.~(\ref{eq:pdferra}, \ref{eq:pdferr}) we sum the maximum deviations
on the observable in each of the parameter directions, and hence
retain both maximal sensitivity to the parameters that vary the most
and estimate the range of allowed values of the cross section.  Given
the case in Fig.~\ref{fig:errasm}, we determine that $m_{\wprm} = 500$
GeV has an uncertainty of ${+8.6}{-6.2}\%$ --- half again as large
on the high side as estimated by the CTEQ method.

\subsection{NLO Differential spectra}
\label{sec:diffnlo}

In the course of an experimental analysis it is common practice to
normalize the number of events generated in an event generator by a
``K-factor''.  This K-factor is generally taken to be the ratio of the
NLO to LO inclusive cross section.  It is then assumed that turning on
successive gluon radiation, called showering, in an event generator
reproduces the correct shapes for the NLO distributions.  While this
is a reasonable approximation for the soft radiation that accompanies
the jets, it does not always reproduce the spectrum of additional
well-separated hard jets \cite{Corcella:2000bw,Sjostrand:2001yu}, or
their effect on the primary jets.  Since experimental reconstruction
efficiencies depend on jet energies, it is important to have the most
accurate prediction possible.

There are two main benefits of looking at the fully differential cross
section.  The first is that this provides a check on the jet
distributions that come from the event generators.  The second is that
we see immediately whether the kinematic regions where perturbation
theory breaks down are relevant to the measurable range of the
distributions.  The shape of the transverse momentum is especially
important in studies with $b$-tags, since the tagging efficiencies
depend on this variable.

\begin{figure}[tbh]
\epsfxsize=3.29in
\centerline{\epsfbox{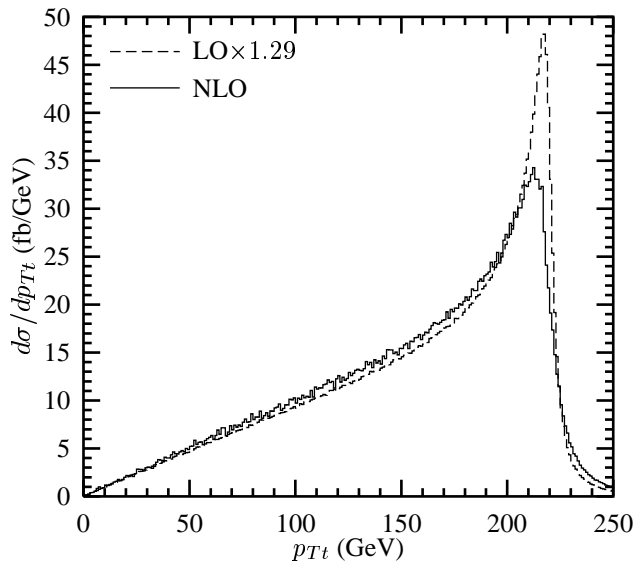}} % 237pt is natural size
\caption{Transverse momentum $p_{T t}$ distribution of the top-quark
at NLO, and LO times a K-factor, for $m_{W^{\prime}}=500$
GeV.\label{fig:ptt}}
\end{figure}

\begin{figure}[tbh]
\epsfxsize=3.29in
\centerline{\epsfbox{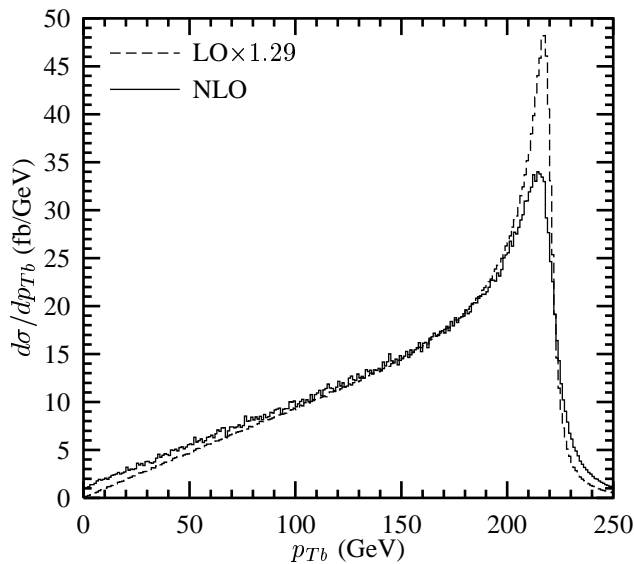}} % 237pt is natural size
\caption{Transverse momentum $p_{T b}$ distribution of the $b$-jet at
NLO, and LO times a K-factor, for $m_{W^{\prime}}=500$
GeV.\label{fig:ptb}}
\end{figure}

In Figs.~\ref{fig:ptt} and \ref{fig:ptb} we show the transverse
momentum ($p_T$) distributions for the top-quark and bottom-jet,
respectively, for $p\bar p\to \wprmp\!\to t\bar b$, and
$m_{\wprm}=500$~GeV at run I of the Tevatron.  For comparison we plot
both the NLO distributions and the LO distributions times a NLO
K-factor of $1.29$.  The $b$-jet is reconstructed using a $k_T$
cluster algorithm~\cite{Ellis:tq} to provide an infrared-safe way of
combining partons.  We use a $k_T$ cone size of $\Delta R = 1$,
similar to a fixed cone size of $0.7$.

\begin{figure}[tbh]
\epsfxsize=3.22in
\centerline{\epsfbox{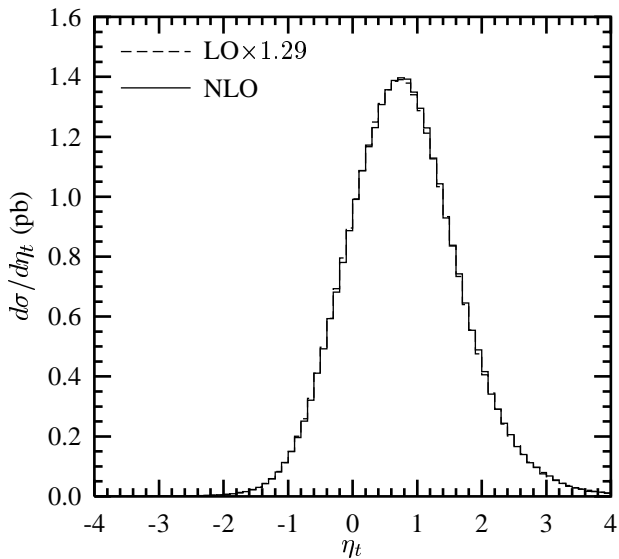}} % 232pt is natural size
\caption{Pseudorapidity $\eta_t$ distribution of the top-quark at NLO, and
LO times a K-factor, for $m_{W^{\prime}}=500$ GeV.\label{fig:etat}}
\end{figure}

\begin{figure}[tbh]
\epsfxsize=3.22in
\centerline{\epsfbox{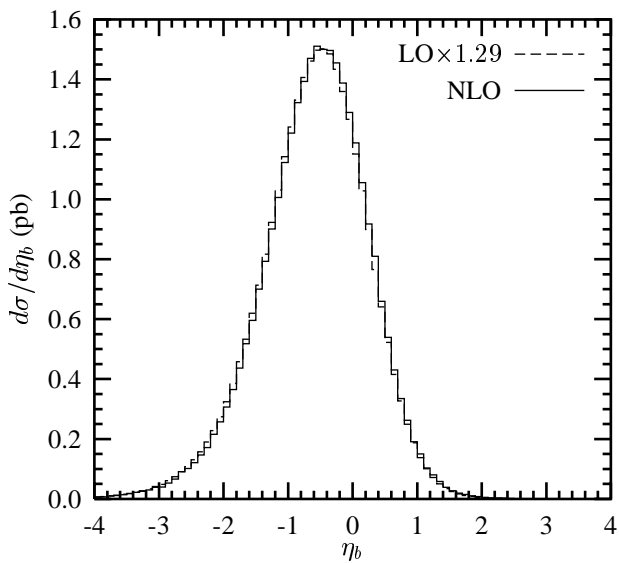}} % 232pt is natural size
\caption{Pseudorapidity $\eta_b$ distribution of the $b$-jet at NLO, and
LO times a K-factor, for $m_{W^{\prime}}=500$ GeV.\label{fig:etab}}
\end{figure}

As is generally the case, the $p_T$ spectra of the $b$-jet and
top-quark are somewhat softened at NLO.  In the top-quark and $b$-jet
$p_T$ distributions in Figs.~\ref{fig:ptt} and \ref{fig:ptb} there is
a much improved behavior near the \wpr resonance at NLO.  Aside from
going to higher orders, it may be possible to further improve the
shape in the resonance region by replacing the Breit-Wigner with a
more dynamical form \cite{Seymour:1995qg}.  While there may be minor
deviations at small transverse momentum due our choice of a massless
bottom quark, the cross section vanishes at low $p_{Tb}$, and $b$-jet
cannot be measured below about 10 GeV, so the effect will not be
visible in the experiments.  The pseudorapidity distributions of the
top-quark and $b$-jet are shown in Figs.~\ref{fig:etat} and
\ref{fig:etab}, respectively.  The shapes of the distributions are so
similar, that the LO curves are completely hidden by the NLO curves.

\begin{figure}[tbh]
\epsfxsize=3.29in
\centerline{\epsfbox{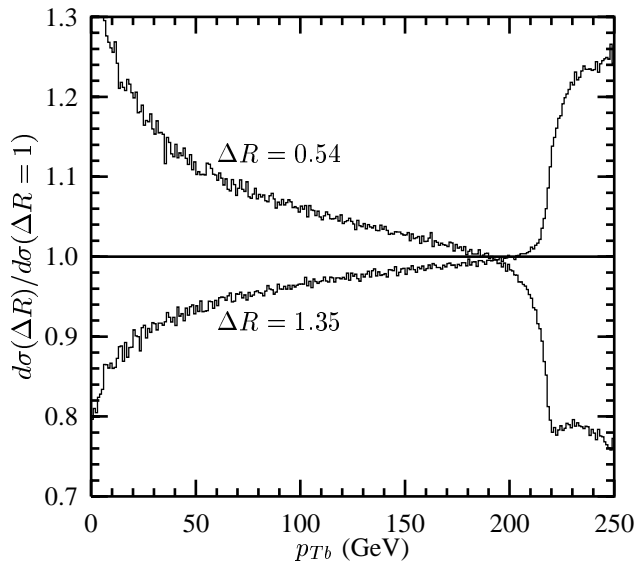}} % 237pt is natural size
\caption{Ratio of distributions of the transverse momentum of the
$b$-jet using different $k_T$ cone sizes at NLO.\label{fig:rb}}
\end{figure}

An important consideration is the effect of the choice of cone size on
the shape of the transverse momentum distributions.  In
Fig.~\ref{fig:rb}, we show the ratio of the $p_T$ distributions of the
$b$-jet for two common choices of $\Delta R=0.54$ and $1.35$ (which
are similar to fixed cones of size $0.4$ and $1.0$, respectively) to
our default choice of $1.0$.  When comparing distributions from event
generators with NLO calculations, we must use the same jet definition.
Otherwise, there can be a systematic shift in the shape of the
distributions of $10$--$20\%$.  This size of this effect is
potentially larger than the effect of the jet-energy resolution on the
reconstruction of jet shapes.

\begin{figure}[tbh]
\epsfxsize=3.29in
\centerline{\epsfbox{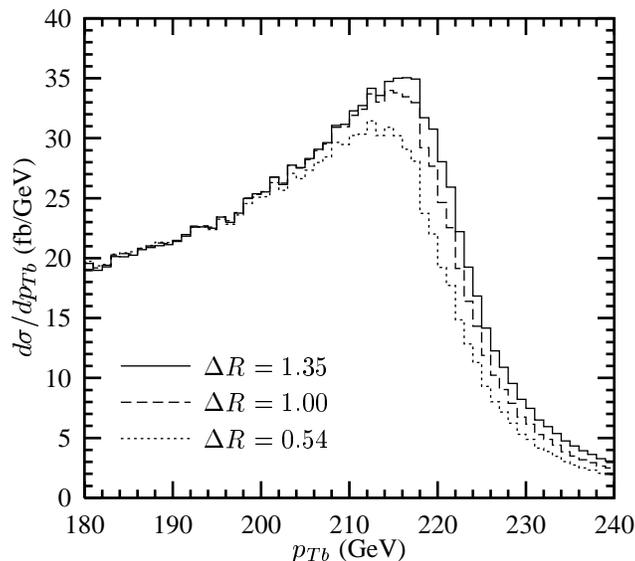}} % 237pt is natural size
\caption{Transverse momentum $p_{T b}$ distribution of the $b$-jet at
NLO using three different cone sizes near the \wpr resonance region
with $m_{W^{\prime}}=500$~GeV.\label{fig:rptb}}
\end{figure}

The overall effect of the choice of cone size is only important for
\wpr production right near resonance.  The cross section is fairly
small at lower transverse momentum, and thus changes in the shape in
that region are not relevant.  The effect at low $p_T$ is further
suppressed in an experiment, because $b$-tagging efficiency tends to
be smaller at low $p_T$ as well.  In Fig.~\ref{fig:rptb} we show the
effect of different cone sizes on the $p_{Tb}$ spectrum near the its
peak at $\sim 215$~GeV, which corresponds to a 500~GeV \wpr resonance.
Experimental analyses will concentrate on this resonant region.
Hence, a mismatch of jet definitions could cause a systematic mistake
in the comparison to the theoretical cross section of as much as $\pm
20\%$.

\begin{figure}[tbh]
\epsfxsize=3.29in
\centerline{\epsfbox{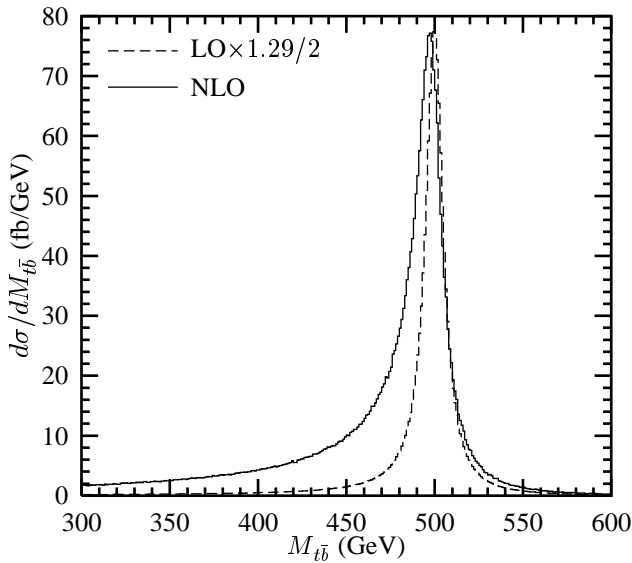}} % 237pt is natural size
\caption{Invariant mass distribution $M_{t\bar b}$ at NLO, and LO
times a K-factor, for $m_{W^{\prime}}=500$~GeV.\label{fig:mtb}}
\end{figure}

While the individual distributions for the \wpr boson are important,
in an experimental analysis we want to fit for the mass of the \wpr
boson.  A likely strategy to find the \wpr is to look for a peak in
invariant mass distribution $M_{t\bar b}$ of the
top-quark/antibottom-jet pair.  We do not perform a full analysis
here, but simply show in Fig.~\ref{fig:mtb} the $M_{t\bar b}$
distributions for $m_{W^{\prime}} = 500$~GeV.  The LO cross section is
shown times the standard K-factor, but is divided by 2 to fit on the
figure.  While the central value of the peak has not shifted by much,
there is a very large tail below the mass of the \wpr boson.  Roughly
half of the cross section is below the peak predicted at LO.  Hard
radiation at NLO has a much larger effect on the correlated
distributions than we might naively expect from the $4\%$ increase in
the width.  Almost $2/3$ of the LO distribution falls in a mass window
of $500\pm 10$ GeV, but only $1/2$ of the NLO distribution falls in a
mass window twice as wide, $500\pm 20$ GeV.  Hence, when considering
the effectiveness of $M_{t\bar b}$, the reconstructible signal over
background may be a factor of 2 smaller than predicted with a LO
calculation.

The two distinguishing features of the \wpr cross section are that the
$b$-jet and top-quark are each at a much larger transverse momentum
than the dominant backgrounds, and there is a peak in the
top-quark/antibottom-jet invariant mass distribution.  While an
optimized phenomenological analysis of the $s$-channel
single-top-quark production cross section at the level of
Ref.~\cite{Stelzer:1998ni} is beyond the scope of this paper, we
estimate the reach in \wpr mass using run I data in the following way:
We begin with the recent limit on the $s$-channel single-top-quark
cross section of $\sim 17$~pb \cite{Abazov:2001ns,Acosta:2001un}.
Assuming that the \wpr boson adds directly to the standard model cross
section, this immediately places a competitive bound of $m_{\wprm}
\agt 380$--$410$~GeV (depending on the allowed decay modes).  To
improve this bound, we note that the invariant mass $M_{Wb\bar b}$ of
the dominant $Wb\bar b$ background ($B$) scales like $1/M^4_{Wb\bar
b}$.  Hence, by reconstructing the \wpr invariant mass, and scaling
the cross section limit by $\sqrt{B}$, we predict a $95\%$
confidence-level limit of $m_{\wprm} \agt 525$--550~GeV should be
attainable in a dedicated analysis of run I data.  In run II of the
Tevatron the $s$-channel single-top-quark cross section will be
measured to $\pm 20\%$ with 2~fb$^{-1}$ of integrated luminosity
\cite{Stelzer:1998ni}.  Given the very low backgrounds at large
invariant masses, limits of 800--900~GeV may be reachable for \wpr
bosons with standard-model size couplings.

\section{Conclusions}

We present a fully differential calculation of \wpr boson production
and decay at next-to-leading order in QCD.  We demonstrate that the
couplings of the \wpr to fermions factorize through NLO in the width,
and in the complete two-to-two cross section.  Any model that contains
a new charged-current gauge particle, with arbitrary vector and
axial-vector couplings, is described by the analytic results of
Secs.~\ref{sec:width} and \ref{sec:sigma}.  In particular, the
completely differential NLO cross section may be used to predict jet
distributions for a given model in terms of $|V^{\prime}_{f_if_j}|$,
$R_t$, and $R_u$.  In most models, left-right mixing is either highly
suppressed, or identically zero, which leads to a value of $R_u=0$.
Hence, by calculating $R_t$, we may translate the results for right-
or left-handed \wpr bosons, appearing in Secs.~\ref{sec:tottop} and
\ref{sec:diffnlo}, into predictions for these models as well.

We use our calculation to estimate the effect of a left- or
right-handed \wpr boson on the single-top-quark cross section.  We
show that the dominant uncertainties in the theoretical prediction
come from the top-quark mass, higher-order QCD, and the parton
distribution functions.  In order to determine the PDF uncertainties,
we propose the use of an improved modified tolerance method (MTM) that
attains maximal sensitivity to the variance of the PDFs, and allows us
to predict asymmetrical uncertainties.  We present the NLO transverse
momentum and pseudorapidity distributions of the top-quark and
bottom-jet.  We describe the effect of jet definitions on these
distributions, and point out the strong sensitivity near the \wpr
resonance region.  We show that the correlated effect of additional
hard radiation on the reconstruction of the \wpr mass peak can be much
greater than naively expected from the increase in the \wpr width.

We predict that the direct search limit on the mass of the \wpr boson
may be improved by using data from runs I and II of the Fermilab
Tevatron.  By reconstructing the top-quark/bottom-jet invariant mass
$M_{tb}$ in the single-top-quark sample, a run I $95\%$
confidence-level limit of $m_{\wprm} \agt 525$--550~GeV should be
attainable for standard-model-like couplings.  At the end of run II,
this limit may be pushed to 800--900~GeV.  The signal at the LHC
should be large enough to reach limits of at least 3--4~TeV.  While
these estimations are based on the known scaling of the backgrounds, a
dedicated phenomenological analysis would be useful.

We conclude by applying our results to one specific example.  In the
top-flavor model of Ref.~\cite{He:1999vp}, the left-handed coupling in
the $\wprm$-$t$-$b$ vertex may be written as
\beq
g_L\sin\zeta\, V^L_{tb} = g\tan\phi\, V_{tb} \,.
\eeq
The coupling to leptons, or the first and second generation of quarks,
can be written
\beq
g_L\sin\zeta\, V^L_{f_if_j} = g\cot\phi\, V_{f_if_j} \,,
\eeq
where $g$ is the standard model SU(2)$_L$ coupling, and $V_{f_if_j}$
is the CKM matrix or diagonal, for quarks or leptons, respectively.
For masses of around 500~GeV, the angle $\phi$ is restricted to be
small \cite{Malkawi:1999sa}, $\sin^2\!\phi < 0.05$.  If we choose
$\sin^2\!\phi = 0.05$, the branching ratio into top quarks is $99\%$;
virtually all of the decays of the \wpr will be into top quarks.  The
enhancement in branching fraction is exactly compensated by a decrease
in production luminosity.  The single-top-quark cross section has $R_t
= \tan^2\!\phi\, \cot^2\!\phi\, R_t^{\mathrm SM}$.  Thus, the direct
search limit for this top-flavor model will be identical to the one
derived with standard-model couplings.  If we wish to set a limit
using a smaller value for $\sin^2\!\phi$, the \wpr width will be the
limiting factor in the mass reconstruction, and we should reevaluate
the jet distributions in that scenario.  By calculating
$|V^{\prime}_{f_if_j}|$, $R_t$, and $R_u$, and using our
next-to-leading order results, we may now accurately compare any model
containing a \wpr boson to direct experimental searches.

\section{Acknowledgements}

The author would like to thank S.~Nandi and T.~Tait for discussions
regarding theoretical models, and P.~Savard for experimental
motivation.  This work is supported by the U.~S.~Department of Energy
under contract No.~DE-AC02-76CH03000.

\end{document}